\documentclass[final,5p,times,twocolumn,unknownkeysallowed]{elsarticle}

\usepackage{doi}
\usepackage{hyperref}
\hypersetup{
	colorlinks=true,        
	linkcolor=blue,         
	citecolor=cyan,         
}

\usepackage[utf8x]{inputenc}
\DeclareUnicodeCharacter{2212}{\textendash}
\usepackage{graphicx}
\usepackage{amssymb}
\usepackage{amsmath}
\usepackage{amsthm}
\usepackage{lineno}
\usepackage{hyperref}
\usepackage{multirow}
\usepackage{bigints}

\usepackage{dcolumn}
\usepackage{bm}
\usepackage{color}

\hypersetup{colorlinks=true,linkcolor=cyan,citecolor=cyan,urlcolor=blue,bookmarks=true}\modulolinenumbers[200]

\biboptions{comma,sort&compress}

\def\a{\alpha}
\def\r{\rho}
\def\s{\sigma}
\def\t{\tau}
\def\m{\mu}
\def\n{\nu}
\def\k{\kappa}
\def\th{\theta}
\def\g{\gamma}\def\G{\Gamma}
\def\L{\Lambda}\def\l{\lambda}
\def\D{\Delta}
\def\la{\langle}
\def\ra{\rangle}
\def\o{\omega}\def\O{\Omega}
\def\d{\delta}
\def\p{\partial}

\newcommand{\be}{\begin{equation}}
\newcommand{\ee}{\end{equation}}
\newcommand{\bea}{\begin{eqnarray}}
\newcommand{\eea}{\end{eqnarray}}

\def\half{\textstyle{\frac{1}{2}}}

\def\bdoc{\begin{document}}
\def\edoc{\end{document}}
\def\beq{\begin{equation}}
\def\eeq{\end{equation}}
\def\bea{\begin{eqnarray}}
\def\eea{\end{eqnarray}}
\def\ben{\begin{enumerate}}
\def\een{\end{enumerate}}
\def\la{\langle}
\def\ra{\rangle}
\def\a{\alpha}
\def\b{\beta}
\def\g{\gamma}
\def\G{\Gamma}
\def\d{\delta}
\def\D{\Delta}
\def\e{\epsilon}

\def\th{\theta}
\def\k{\kappa}
\def\l{\lambda}
\def\m{\mu}
\def\n{\nu}
\def\o{\omega}
\def\p{\pi}
\def\r{\rho}
\def\s{\sigma}
\def\t{\tau}
\def\L{{\cal L}}
\def\S{\Sigma }
\def\gsim{\; \raisebox{-.8ex}{$\stackrel{\textstyle >}{\sim}$}\;}
\def\lsim{\; \raisebox{-.8ex}{$\stackrel{\textstyle <}{\sim}$}\;}
\def\gtrsim{\gsim}
\def\lessim{\lsim}
\def\loc{{\rm local}}
\def\vm{v_{\rm max}}
\def\bh{\bar{h}}
\def\del{\partial}
\def\nab{\nabla}
\def\half{{\textstyle{\frac{1}{2}}}}
\def\fourth{{\textstyle{\frac{1}{4}}}}

\def\bD{{\bf D}}
\def\bE{{\bf E}}
\def\bF{{\bf F}}
\def\bB{{\bf B}}
\def\bP{{\bf P}}
\def\bV{{\bf v}}
\def\bv{{\bf v}}
\def\bx{{\bf x}}
\def\by{{\bf y}}
\def\bz{{\bf z}}
\def\ba{{\bf a}}
\def\bd{{\bf d}}
\def\bs{{\bf s}}
\def\bn{{\bf n}}
\def\bp{{\bf p}}

\def\O{\Omega}

\def\br{{\bf r}}
\def\bnab{{\bf \nab}}

\def\tE{\tilde{E}}
\def\tL{\tilde{L}}

\journal{Journal Name}

\begin{document}

\begin{frontmatter}

\title{Circular orbits around higher dimensional Einstein and pure Gauss-Bonnet rotating black holes
 }

\author[mainaddress1]
{Naresh Dadhich}
\ead{nkd@iucaa.in}
\author[mainaddress2,mainaddress3,mainaddress4,mainaddress5,mainaddress6]
{Sanjar Shaymatov\cortext[cor2]{Corresponding author}\corref{cor2}}
\ead{sanjar@astrin.uz}

\address[mainaddress1]{Inter University Centre for Astronomy \&
Astrophysics, Post Bag 4, Pune 411007, India}
\address[mainaddress2]{Institute for Theoretical Physics and Cosmology, Zheijiang University of Technology, Hangzhou 310023, China}
\address[mainaddress3]{Akfa University,  Milliy Bog Street 264, Tashkent 111221, Uzbekistan}
\address[mainaddress4]{Ulugh Beg Astronomical Institute, Astronomicheskaya
33, Tashkent 100052, Uzbekistan}
\address[mainaddress5]{Institute of Fundamental and Applied Research, National Research University TIIAME, Kori Niyoziy 39, Tashkent 100000, Uzbekistan}
\address[mainaddress6]{National University of Uzbekistan, Tashkent 100174, Uzbekistan}

\date{Received: date / Accepted: date}

\begin{abstract}
In this paper we study circular orbits around higher dimensional rotating Myers-Perry and pure Gauss-Bonnet (GB) black holes. It turns out that for the former there occurs no potential well to harbour bound and thereby stable circular orbits. The only circular orbits that could occur are all unstable and their radius is bounded from the below by that of the photon circular orbit. On the other hand bound and stable circular orbits do exist for pure GB/Lovelock rotating black holes (the metric is though not an exact solution of pure Lovelock vacuum equation but it satisfies the equation in the leading order and has all the desired properties) in dimensions, $2N+2 \leq D \leq 4N$ (for $N=2$ pure GB in $D = 6, 7, 8$) where $N$ is the degree of Lovelock polynomial. Thus bound and stable circular orbits could exist around higher dimensional rotating black holes only for pure GB/Lovelock gravity. This property is a nice discriminator between Myers-Perry and pure GB/Lovelock rotating black holes.

\end{abstract}

\begin{keyword}
Circular orbits\sep Einstein and Gauss-Bonnet black holes
\end{keyword}

\end{frontmatter}

\linenumbers

\section{Introduction}
\label{introduction}

Black holes are formed by gravitational collapse -- accretion process. Accreting matter if it has non-zero angular momentum, it would encounter centrifugal potential barrier. Unlike Newtonian theory, in general relativity (GR) there exists a threshold limit on angular momentum below which particles encounter no barrier and they could fall in positing angular momentum onto the central object.  This threshold limit is given by the angular momentum of the innermost stable circular orbit (ISCO). Thus existence of ISCO becomes the critical necessary condition for transmitting angular momentum to the central object. That would therefore play the crucial determining role for formation of rotating black hole by gravitational collapse -- accretion process.

Thus the question arises -- does ISCO always exist in black hole spacetimes? For existence of stable circular orbit; i.e. occurrence of minimum of effective potential giving rise to potential well is required to harbour bound orbits. It is well known that bound orbits around a static object in GR exists only in four dimension and none else \cite{Dadhich13}. Therefore there cannot occur bound orbits in higher dimensions and thereby there occurs no angular momentum threshold for carrying angular momentum down to black hole. It turns out that this feature of non-occurrence of bound orbits/ISCO is also carried over to higher dimensional Myers-Perry rotating black hole.

Unlike GR, bound/ISCO orbits do exist for pure Lovelock\footnote{Pure Lovelock means the Lovelock Lagrangian and the equation of motion have single $N$th order term without sum over lower orders~\cite{Dadhich12}. Here $N$ is the degree of homogeneous Riemann curvature polynomial in Lovelock action.} static black holes in dimensions, $2N+2 \leq D \leq 4N$; i.e. for pure GB in dimensions, $D= 6, 7, 8$~\cite{Dadhich13}. ISCO would define the  threshold limit for angular momentum for particle to carry spin to black hole. That means an accretion process could be set in for pure Lovelock black holes for obtaining a rotating black hole. In pure GB/Lovelock gravity there does not exist an exact solution, like the Myers-Perry solution, of the vacuum equation. There does however exist a metric conjured for pure GB rotating black hole in Ref.~\cite{Dadhich-Ghosh13}, which is obtained following the procedure of Ref.~\cite{Dadhich13b} by which the Kerr metric was obtained without solving the field equations.  Though it is not an exact solution, yet it has all the desired properties of a rotating black hole, and satisfies the equation in the leading order.

For probing the questions raised above, we shall in this paper study circular orbits for Myers-Perry and pure GB rotating black holes in higher dimensions. Timelike and null geodesics around five dimensional rotating black hole were studied \cite{Frolov-Stojkovic03} and it was shown that there cannot exist stable circular orbits. A complete analysis and characterization of constants of motion in higher dimensional rotating black holes were carried out in \cite{Page07,Krtous07}.

The paper is organized as follows: In Sec.~\ref{sec:metric} we briefly recall Myers-Perry rotating black hole metric which is followed by particle dynamics and effective potential in Secs.~\ref{sec:dynamics} and \ref{sec:MP}. We discuss about physical motivation for pure Lovelock theory in Sec.~\ref{sec:PLgravity}. Sec.~\ref{sec:GB} we devote to pure GB rotating black hole and orbits around it. We discuss occurrence of one or two horizons in Sec.~\ref{sec:one or two}. We end up with discussion and conclusion in Sec.~\ref{sec:Conclusion}. Throughout we use a system of units in which gravitational constant and velocity of light are set to unity.

\section{Higher dimensional rotating black hole }\label{sec:metric}

There is the well known Myers-Perry solution \cite{Myers-Perry86} describing a rotating black hole in higher dimensions. The line element for that is given by
\textcolor{black}{\begin{eqnarray}\label{Eq:D}
ds^2&=&-dt^2+r^2d\beta^2 + \sum_{i=1}^{n}(r^2+a^2_{i})\left(d\mu_{i}^2+\mu_{i}^2d\phi^2_{i}\right)\nonumber\\&+&\frac{\mu r}{\Pi F}\left(dt +\sum_{i=1}^{n}a_{i}\mu_{i}^2d\phi_{i}\right)^2 +\frac{\Pi F}{\Delta}dr^2\, ,
\end{eqnarray}
with
\begin{eqnarray}\label{Eq:D1}
F &=& 1-\sum_{i=1}^{n}\frac{a_{i}^2\mu_{i}^2} {r^2+a_{i}^2}\, , \nonumber\\
\Pi &=&\prod_{i=1}^{n}(r^2+a_i^2)
\, , \nonumber\\
\Delta &=& \Pi -2\mu r^{2n-D+3} \, .
\end{eqnarray}
Here $\mu$ and $a_{i}$ are black hole mass and rotation parameters, and $\mu_i$ and $\beta$ are related by the following expressions (see, Ref.~\cite{Myers2011}), 
\begin{eqnarray}\label{Eq:2n+2}
\sum_{i=1}^{n} \mu_i^2 + \beta^2 &=& 1\, ,\\
\label{Eq:2n+1}
\sum_{i=1}^{n} \mu_i^2 &=& 1\, ,
\end{eqnarray}
for $D=2n+2, 2n+1$ respectively, the latter results when $\beta=0$ is satisfied in the former as well as in the metric. Note that $\mu_i$ are the direction cosines, for example, $\mu_1$ and  $\mu_2$ for $D=5,6$ dimensions will respectively read as 
\begin{eqnarray}
\mu_1=\sin\theta\, \mbox{~~and~~}\mu_2=\cos\theta\, ,
\end{eqnarray}
and 
\begin{eqnarray}
\mu_1=\sin\theta\, ,\,\,  
\mu_2=\cos\theta\sin\chi\, \,\,\mbox{and} \,\,\,
\beta=\cos\theta\cos\chi\, .
\end{eqnarray}} 
Note that in higher dimensions, black hole can have more than one rotations, and $n=[(D-1)/2]$ is the maximum number of rotations it can have in the given dimension $D$; i.e., $n=2$ for $D=5,6$ dimensions. 

The horizons of black hole are located at the real positive roots of $\Delta=0$ which for $n=1$ in $D=5$ would read as~{\cite{Shaymatov19a}}
\begin{eqnarray}
r^2 + a^2 -2\mu = 0\, .
\end{eqnarray}
This gives $r_h = \sqrt{2\mu-a^2}$ giving the extremal limit, $a=\sqrt{2\mu}$ {for which horizon coincides with the origin $r=0$}.

\begin{figure*}
\centering
 \includegraphics[width=0.32\textwidth]{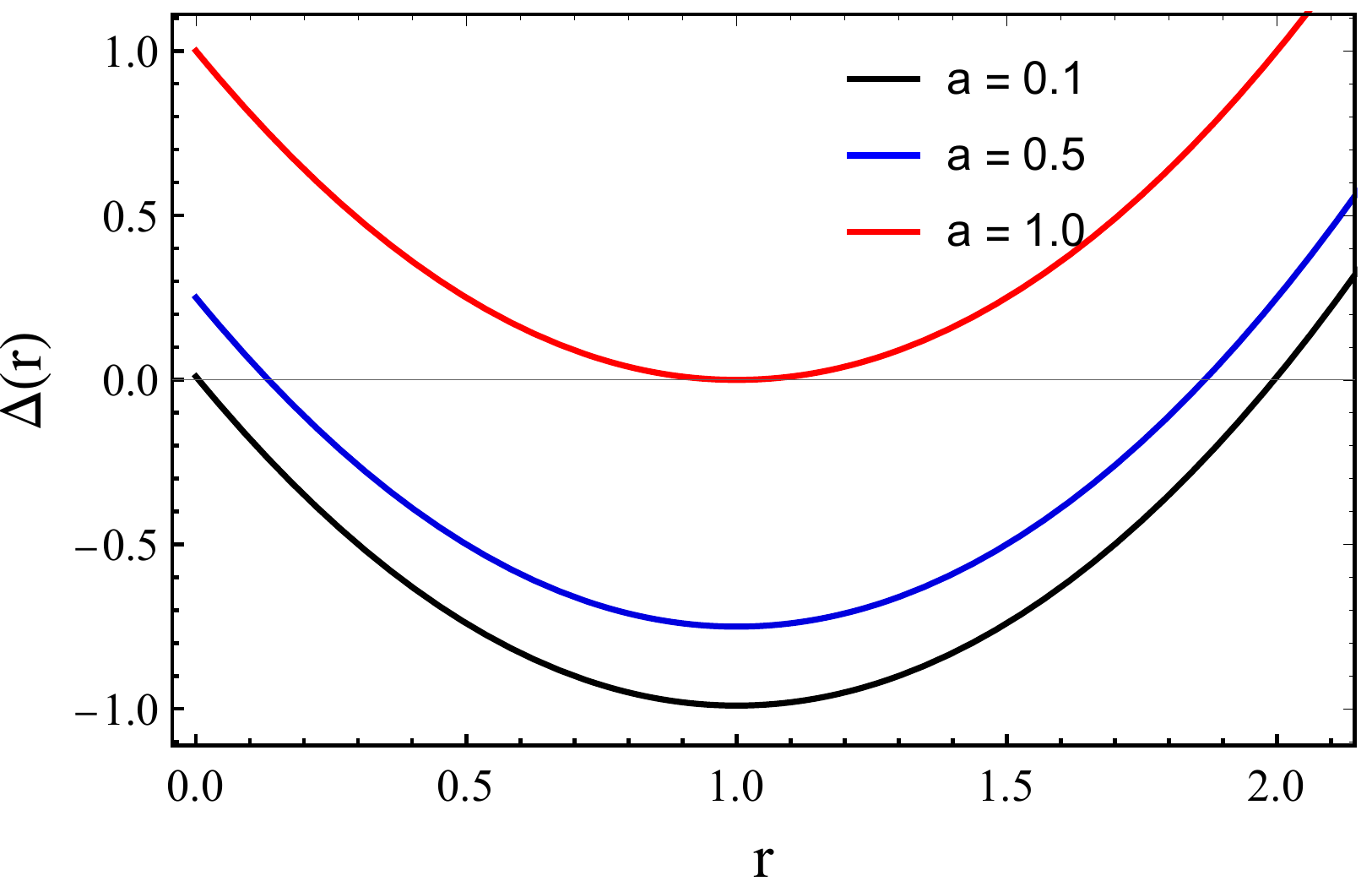}
 \includegraphics[width=0.32\textwidth]{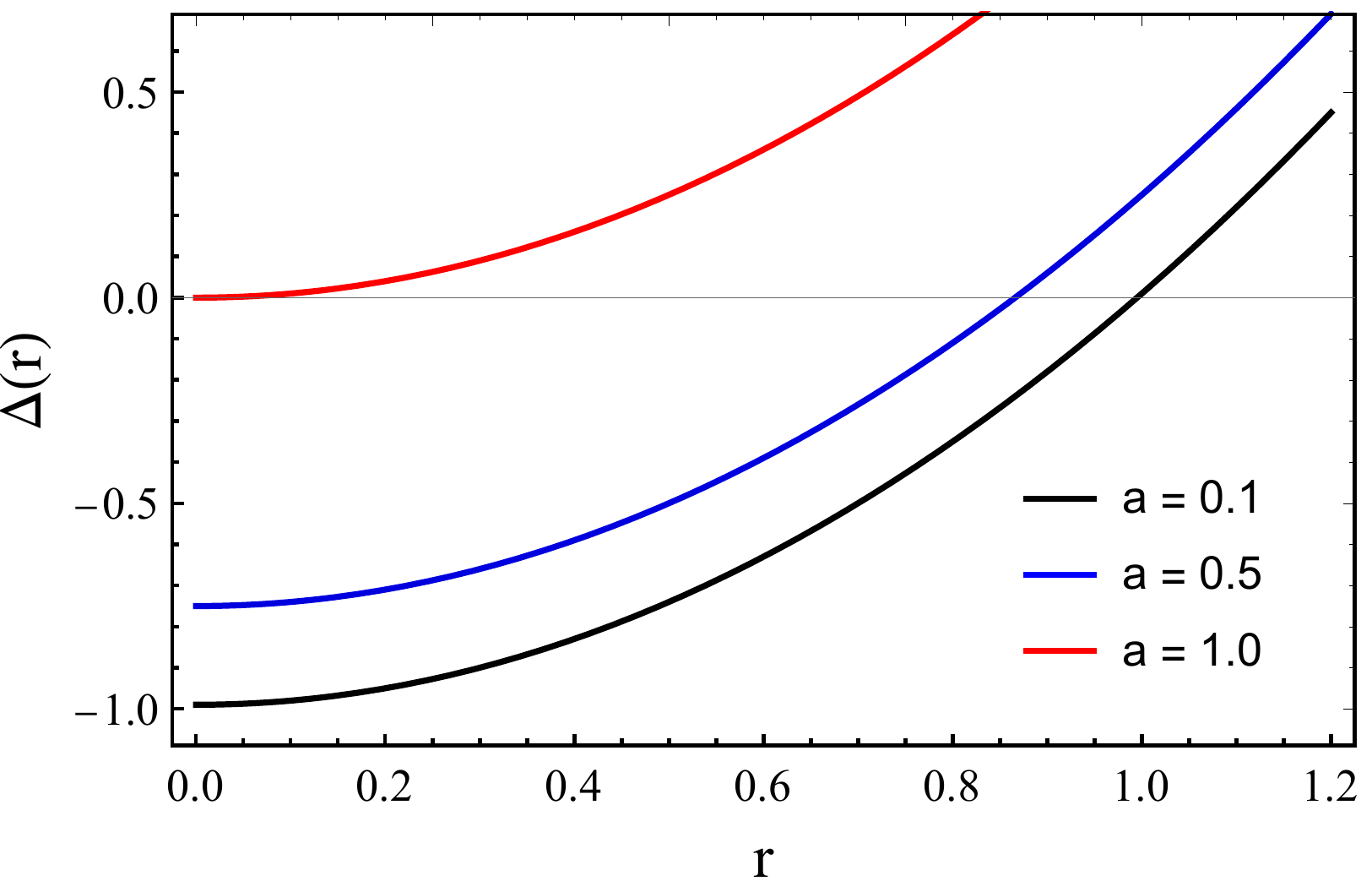}
\includegraphics[width=0.32\textwidth]{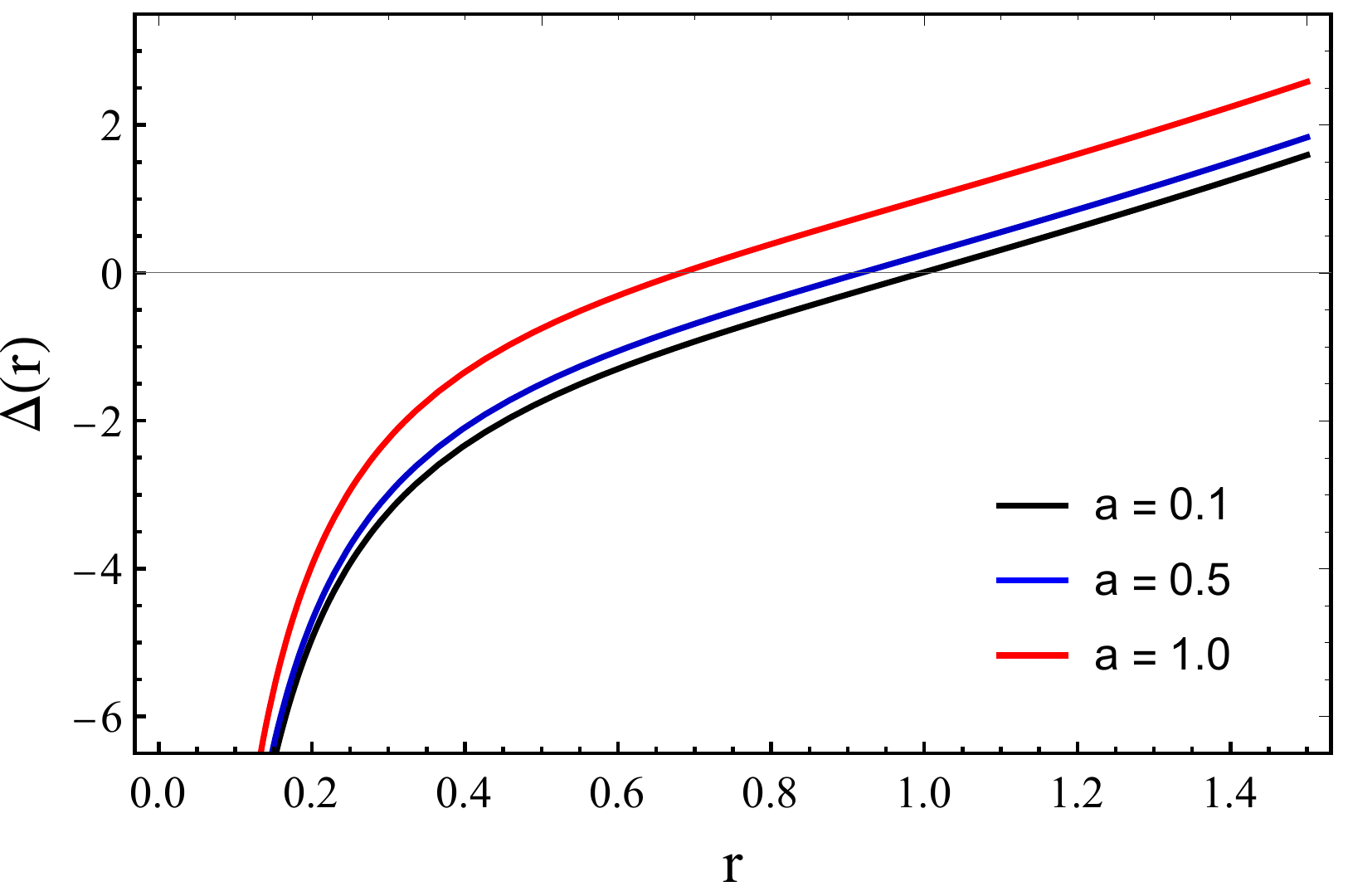}

 \caption{\label{fig:delta1} $\Delta(r)$ is plotted for $D=4,5,6$ in left, middle and right panels respectively. Note that for a single rotation, two horizons occur only in four dimension and not in any higher dimension. }
\end{figure*}

\begin{figure*}
\centering
  \includegraphics[width=0.45\textwidth]{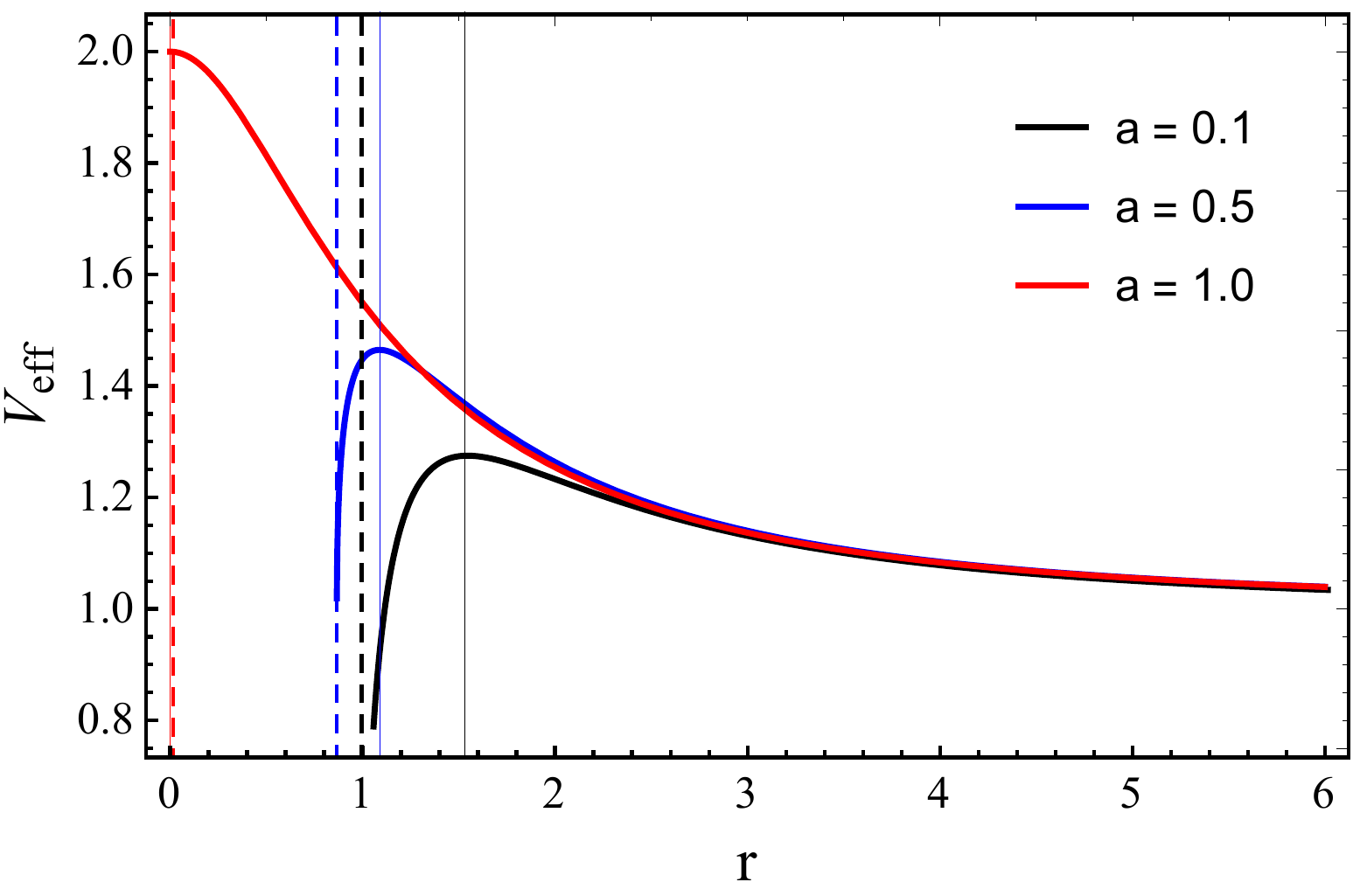}
   \includegraphics[width=0.45\textwidth]{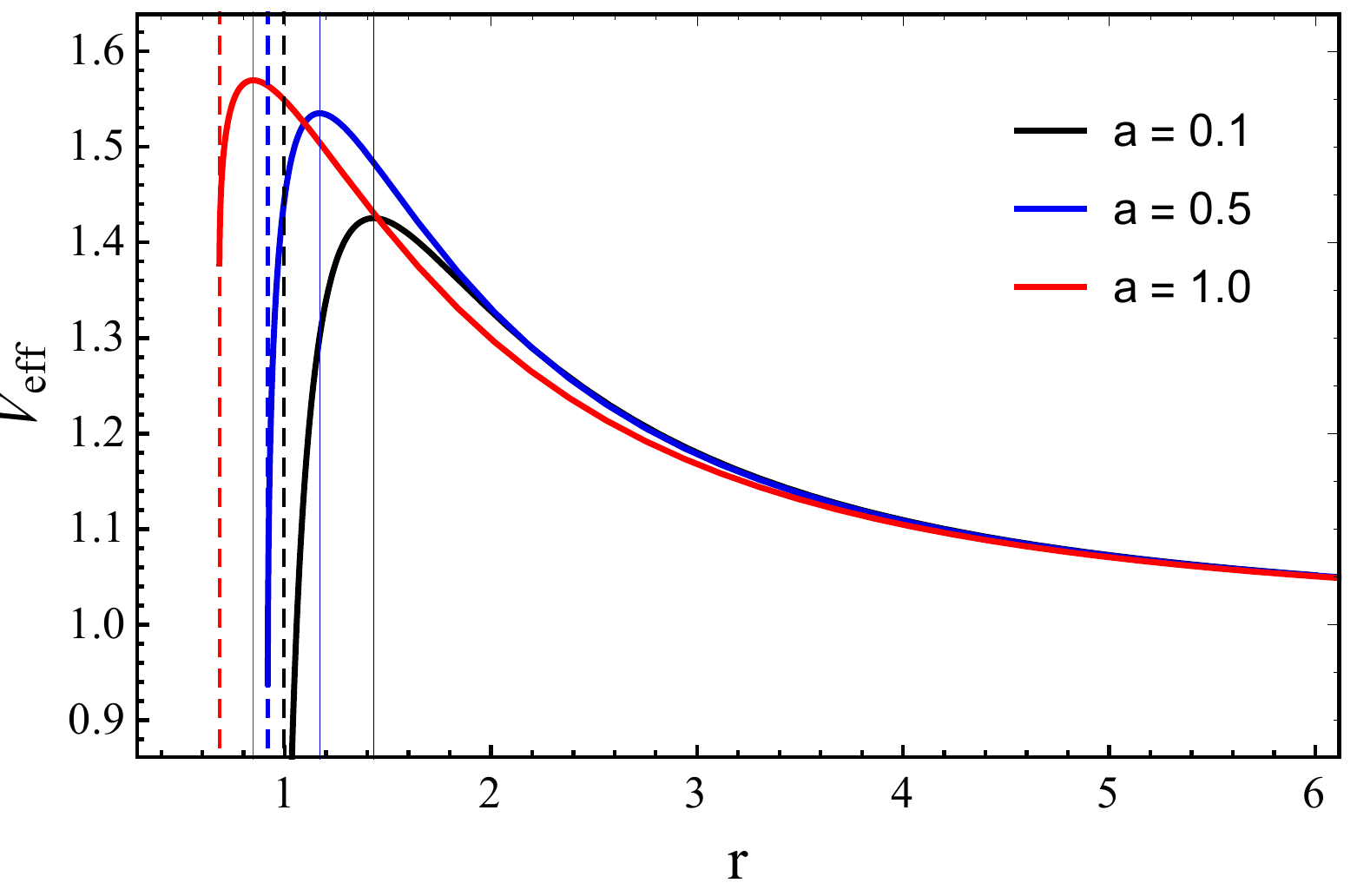}
\caption{\label{fig_eff} $V_{eff}$ for $n=1$ and $\mathcal{L}=2$ in $D=5, 6$ (left/right panels). Vertical dashed lines indicate location of horizon $r_{h}$ while thick lines indicate radius of innermost unstable circular orbit -- the existence threshold. }
\end{figure*}

\begin{figure*}
\centering
  \includegraphics[width=0.45\textwidth]{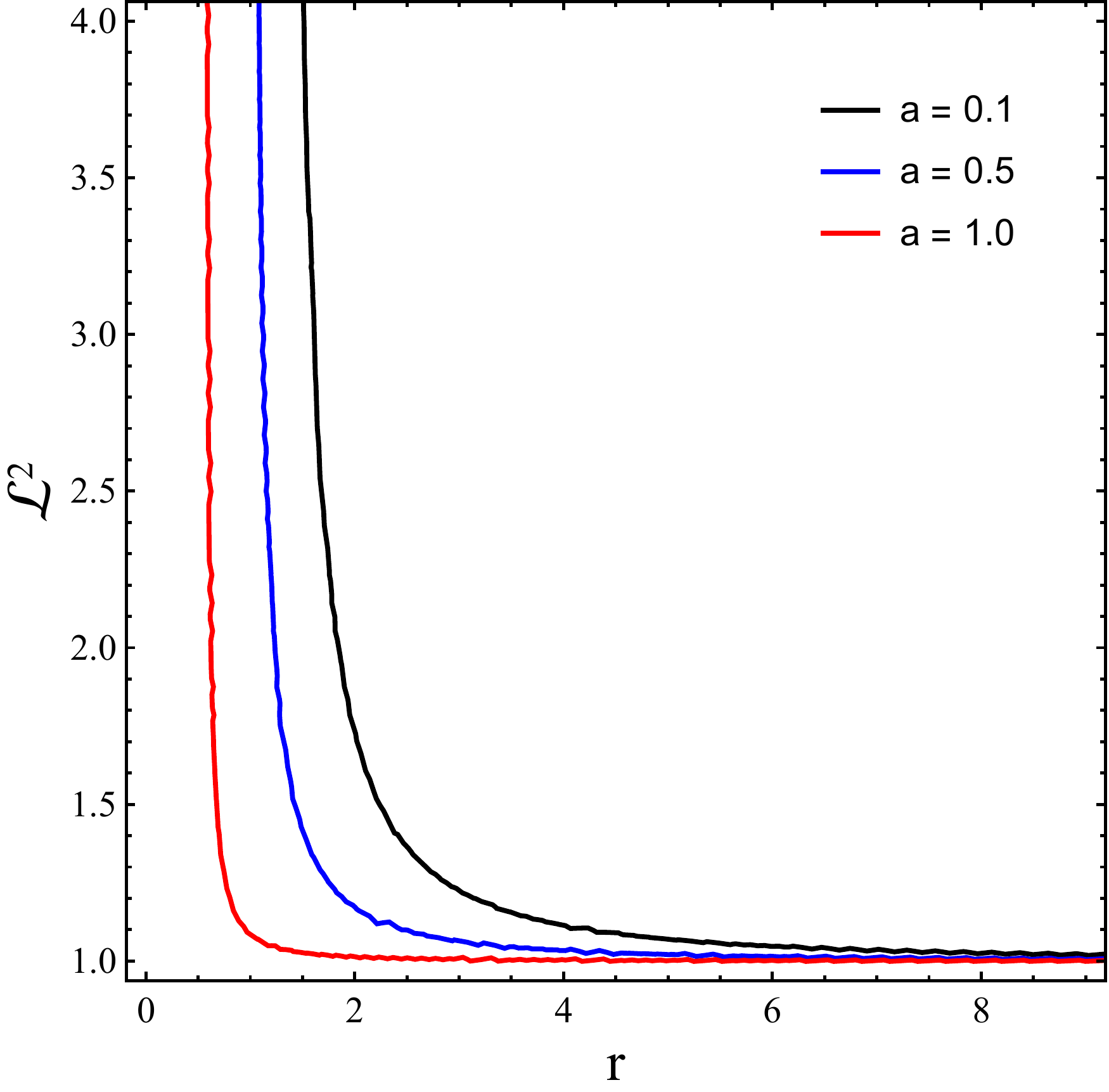}
  \includegraphics[width=0.45\textwidth]{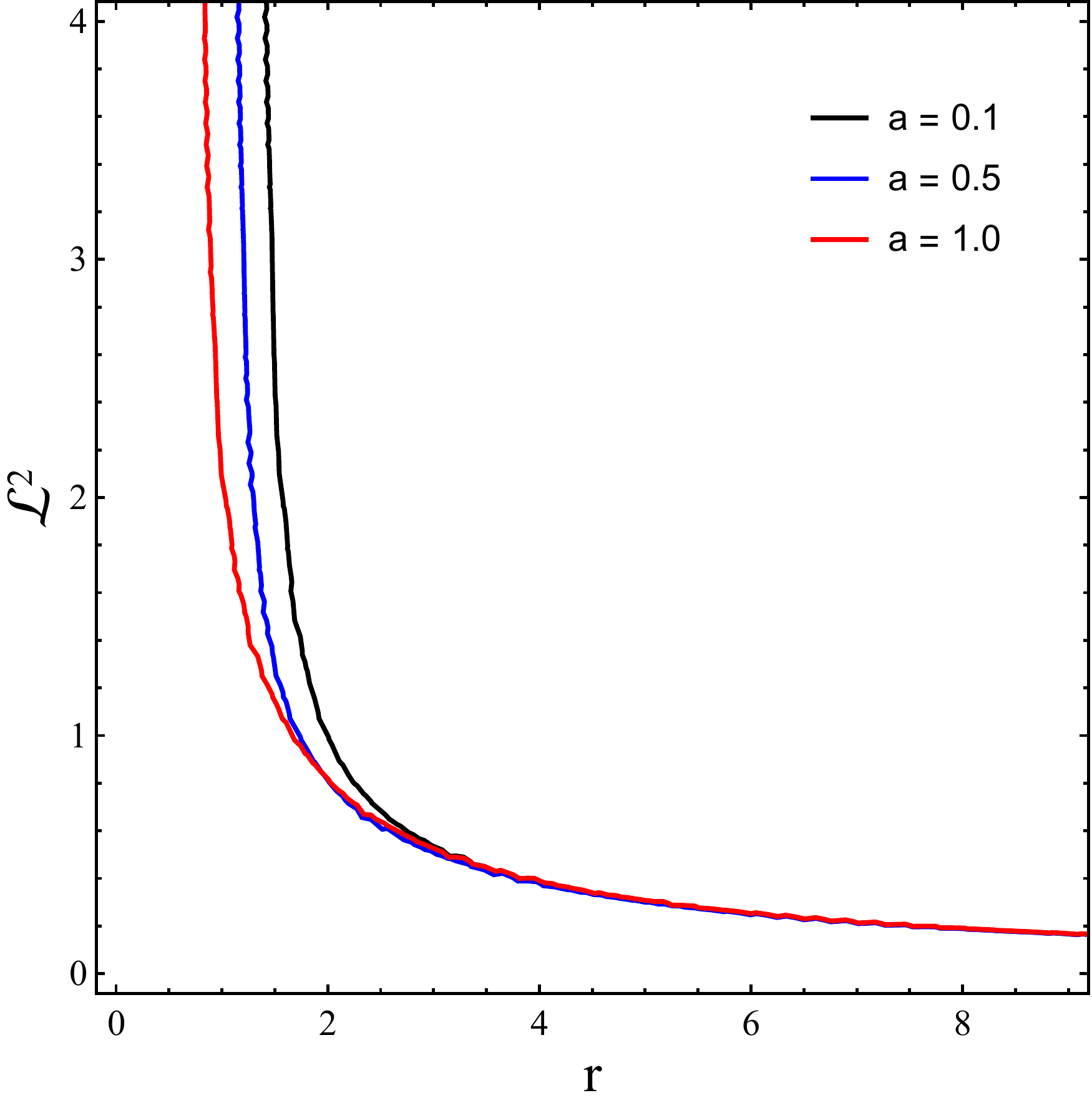}

\caption{\label{fig1} $\mathcal{L}^2$ is against the rotation parameter $a$ in $D=5, 6$ (left/right panels).
}
\end{figure*}

Note that for black hole having single rotation, two horizons occur only in four dimension and not in any higher dimension (see Fig.~\ref{fig:delta1}). That is why there occurs no extremal limit for rotation in higher dimensions for $n=1$. In Figs.~\ref{fig_eff} and \ref{fig1}, $a=1$ {(in all Figs. the mass parameter is set equal to unity)} is not the extremal limit.


\section{Particle dynamics }\label{sec:dynamics}

Now we come to motion of particle of rest mass $m$ around higher dimensional rotating black hole having a single rotation. We begin by writing the standard Hamiltonian for motion,
\begin{eqnarray}
 H  \equiv \frac{1}{2}\,g^{\alpha\beta}\,\pi_\alpha \pi_\beta\, ,
\label{Eq:H}
\end{eqnarray}
with four momentum $\pi^\alpha =mu^{\alpha}$ and the rest mass, {$H=-m^2/2$}~\cite{Misner73}.  Hamilton's equations of motion are then given by
\begin{eqnarray}
  \frac{dx^\alpha}{d\varsigma} &=& \frac{\partial H}{\partial \pi_\alpha}   \, ,
\label{Eq:H-1}  \\
  \frac{d\pi_\alpha}{d\varsigma} &=& - \frac{\partial H}{\partial x^\alpha} \, ,
\label{Eq:H-2}
\end{eqnarray}
where the affine parameter $\varsigma$ is related by proper time  $ {\varsigma}=\tau/m$.  Following Hamilton–Jacobi equation, we write  the action $S$ as
\begin{eqnarray}\label{Eq:action}
S&=& -\frac{1}{2}m^2\tau-Et+L\varphi+S_{r}(r)+S_{\theta}(\theta)\nonumber\\&& + S_{\chi}(\chi)+S_{\psi}(\psi)\, .
\end{eqnarray}
Here the quantities $E \equiv -\pi_t$ and $L \equiv \pi_{\varphi}$ are the conserved quantities of motion, referred to the energy and angular momentum of particle.

Let us then rewrite the Hamiltonian,
\begin{eqnarray}\label{Eq:separable}
-m^2&=& g^{tt}E^2- 2g^{t\phi}E\, L+ g^{\phi\phi}L^2\nonumber\\&+& g^{rr}\pi_{r}^2+g^{\theta\theta}\pi_{\theta}^2 + g^{\chi\chi}\pi_{\chi}^2+ g^{\psi\psi}\pi_{\psi}^2\, .
\end{eqnarray}
Note that the system is described by five independent constants of motion of which we have specified three; i.e. $E$, $L$ and $m^2$. The other two are related to the latitudinal motion which become irrelevant as we restrict motion to the equatorial plane; i.e. $\theta = \pi/2$.

Now the radial equation of motion takes the form
\begin{eqnarray}\label{Eq:limit}
\dot{r}^{2} = \Big(\mathcal{E} -
\mathcal{E_{+}}(r)\Big)\Big(\mathcal{E}
-\mathcal{E_{-}}(r)\Big)\, ,
\end{eqnarray}
where $\mathcal{E_{+}}$ and $\mathcal{E_{-}}$ are the two roots of the quadratic equation $\dot{r} = 0$. Since $\dot{r}^2\geq 0$ always, hence it must be either $\mathcal{E}>\mathcal{E}_{+}(r)$ or $\mathcal{E}<\mathcal{E}_{-}(r)$. However, we choose  $\mathcal{E}_{+}(r) = V_{eff}(r)$ which alone is physically acceptable for four momentum to be future pointing \cite{Atamurotov21JCAP}. For further analysis we shall for simplicity choose that black hole has only one rotation parameter. As our main purpose is to study the qualitative aspects of the particle dynamics for that it would not matter much whether black hole has only one or both rotations. This however does matter in the context of extremality condition of black hole. It turns out that there occurs no extremality condition when number of rotations is less than the maximum allowed; i.e. $<n$,  \cite{Shaymatov21a}. {However, in the marginal case of five dimension there occurs extremality condition for single rotation~\cite{Shaymatov19a}, and it turns out that it could be overspun at linear order perturbations. This result is however overturned when non-linear order perturbations are taken into account~\cite{Shaymatov20a,Shaymatov19c,Shaymatov21d}.}

We thus define the effective potential for equatorial motion of particle in the field of a six dimensional black hole with a single rotation, and it is given by
{
\begin{eqnarray}\label{Veff_gen}
V_{eff}(r)&=&- \frac{g_{t\phi }}{g_{\phi \phi }}\mathcal{L}+\sqrt{\frac{\Delta}{g_{\phi\phi}\,
}\left(1+\frac{\mathcal{L}^2}{g_{\phi\phi}}\right) }\, .
\end{eqnarray}}
Here we have used the specific physical quantities,  $\mathcal{E}=E/m$ and $\mathcal{L}=L/m$ and have set $m^2=1$.

\section{Orbits in $D=5, 6$}\label{sec:MP}

From Eq.~(\ref{Veff_gen}) we write the effective potential $V_{eff}(r)$ for $D=5, 6$ dimensions in the following form:
%
\begin{eqnarray}
V_{eff}^{5D}(r)&=&\frac{a \mu \mathcal{L}}{r^4+\left(r^2+\mu\right)a^2 }\nonumber\\&+&\frac{ r\Big(r^4+\left(r^2+\mu\right)a^2 +r^2 \mathcal{L}^2\Big)^{1/2}}{r^4+ \left(r^2+\mu\right)a^2}\nonumber\\&\times &\left(r^2-\mu+a^2\right)^{1/2}\, ,\\
V_{eff}^{6D}(r)&=&\frac{a \mu \mathcal{L}}{r^5+\left(r^3+\mu\right)a^2 }\nonumber\\&+&\frac{ r\Big(r^5+\left(r^3+\mu\right)a^2 +r^3 \mathcal{L}^2\Big)^{1/2}}{r^5+ \left(r^3+\mu\right)a^2}\nonumber\\&\times &\left(r^3-\mu+a^2 r\right)^{1/2}\, .
\end{eqnarray}
On expanding for large $r$, these take the form
\begin{eqnarray}
V_{eff}^{5D}(r\to r_{\infty})&\sim & 1+\frac{\left(\mathcal{L}^2-\mu\right)}{2r^2}+\mathcal O\left(\frac{1}{r^{4}}\right)\, ,\\
V_{eff}^{6D}(r\to r_{\infty})&\sim &1+\frac{\mathcal{L}^2}{2r^2}-\frac{\mu }{2r^3} +\mathcal O\left(\frac{1}{r^{4}}\right)\, .\nonumber\\
\end{eqnarray}

As shown in Fig.~\ref{fig_eff}, this clearly shows that $V_{eff} \geq 1$ always for the latter while for the former when $\mathcal{L}^2 > \mu$. Further it has only one extremum which is a maximum and there is no minimum. This means there can neither occur any bound orbit nor a stable circular orbit. This is the characteristic feature of particle motion for rotating black holes in higher dimensions.

We now explicitly show that stable circular orbits cannot exist. The conditions for circular orbits are $\dot{r}=0$ and $\ddot{r}=0$ simultaneously. The former defines $V_{eff}$ while the latter requires its derivative to be zero; i.e.$ \frac{\partial V_{eff}(r)}{\partial r}=0$, giving
\begin{eqnarray}\label{Eq:first_der2}
&&\left(a^2-\mu+r^2\right)^{-1/2} \left[a^2 \left(\mu+r^2\right)+r^2 \left(r^2+\mathcal{L}^2\right)\right]^{-1/2}\nonumber\\&\times &\left[a^2 \left(\mu+r^2\right)+r^4\right]^{-1}\left[a^4 \mu+2 a^4 r^2-a^2 \mu^2\right.\nonumber\\&+&\left. r^2 \mathcal{L}^2 \left(2 a^2-2 \mu+3 r^2\right)+6 a^2 r^4-3 \mu r^4+4 r^6\right]\nonumber\\&-&\frac{2 r \left(a^2+2 r^2\right)}{\left[a^2 \left(\mu+r^2\right)+r^4\right]^2}\bigg[a \mu \mathcal{L}+ r \left(a^2-\mu+r^2\right)^{1/2} \nonumber\\&\times &\Big(a^2 \left(\mu+r^2\right)+r^2 \left(r^2+\mathcal{L}^2\right)\Big)^{1/2}\bigg]=0\, ,\\ \nonumber\\
\label{Eq:first_der22}
&&\left(a^2r-\mu+r^3\right)^{-1/2} \left[a^2 \left(\mu+r^3\right)+r^3 \left(r^2+\mathcal{L}^2\right)\right]^{-1/2}\nonumber\\&\times &\frac{1}{2}\left[a^2 \left(\mu+r^3\right)+r^5\right]^{-1}\left[3 a^4 r \left(\mu+2 r^3\right)\right.\nonumber\\&-&  2 b^2 \left(\mu^2-8 r^6-3 r^4 \mathcal{L}^2\right)-7 \mu r^5-5 \mu r^3 \mathcal{L}^2\nonumber\\ &+& \left. 10 r^8+8 r^6 \mathcal{L}^2\right]\nonumber\\&-&\frac{ \left(3a^2r^2+5 r^4\right)}{\left[a^2 \left(\mu+r^3\right)+r^5\right]^2}\bigg[a \mu \mathcal{L}+ r \left(a^2r-\mu+r^3\right)^{1/2} \nonumber\\&\times &\Big(a^2 \left(\mu+r^3\right)+r^3 \left(r^2+\mathcal{L}^2\right)\Big)^{1/2}\bigg]=0\, ,
\end{eqnarray}
for $D=5, 6$ respectively. Eqs.~(\ref{Eq:first_der2}) and (\ref{Eq:first_der22}) have only one positive root for which $V_{eff}$ reaches its maximum (see Fig.~\ref{fig_eff}). However, it is difficult to solve the equations  analytically and hence we have resorted to numerics.

On solving the above equations for angular momentum $\mathcal{L}$ we obtain respectively for five and six dimensions,
\begin{eqnarray}\label{Eq:An1}
\mathcal{L}^2_{\pm{5D}}(r)&=&\frac{\mu\left(r^2\mp a\mu^{1/2}+a^2\right)^2}{r^2\left[(r^2\pm 2a\mu^{1/2}-2\mu\right] }\, , \\
\mathcal{L}^2_{\pm{6D}}(r)&=&\frac{a^4 \left(6 r^4+9 \mu r\right)+2 a^2 \left(6 r^6-\mu r^3-5\mu^2\right)}{\mu^{-1}r^3 \left[\left(5\mu-2 r^3\right)^2-24 a^2 \mu r\right]}\nonumber\\&+&\frac{3 \mu\left(2 r^3-5\mu\right) r^2}{ \left[\left(5\mu-2 r^3\right)^2-24 a^2 \mu r\right]} \nonumber\\&\mp &\frac{2 \sqrt{6} a \sqrt{\mu r} \left(3 a^2+5 r^2\right) \left(a^2 r+r^3-\mu\right)}{\mu^{-1}r^3 \left[\left(5\mu-2 r^3\right)^2-24 a^2 \mu r\right]}\, .\nonumber\\
\label{Eq:An2}
\end{eqnarray}

The radial profiles of the angular momentum of the circular orbits for $D=5,6$ are shown in Fig.~\ref{fig1}.  As $a$ increases, the curves shift toward left to smaller $r$. Minimum of angular momentum identifies the innermost stable circular orbit (ISCO). On expanding Eqs.~(\ref{Eq:An1}) and (\ref{Eq:An2}), it follows that $\mathcal{L}^2 \to \mu, 0$ as $r \to \infty$ respectively for $D=5, 6$. It never attains a minimum, though it tends to finite value $\mu=1$ for the former while it goes to zero for the latter. In five dimension, a stable circular orbit would occur at infinity but it is not the ISCO because there is no minimum for $\mathcal{L}^2$ while for six dimension there occurs no stable orbit.

Another limit on existence of circular geodesics is given by the photon circular geodesics with radius given by divergence of the angular momentum. One can see from the expressions in Eqs. (\ref{Eq:An1}-\ref{Eq:An2}) that would happen only when
\begin{eqnarray}\label{Eq:ph_con}
r^2\pm 2a\mu^{1/2}-2\mu=0\, ,\\
\left(5\mu-2 r^3\right)^2-24 a^2 \mu r=0\, .
\end{eqnarray}
for $D=5,6$ respectively. Here $\pm$ refer to prograde and retrograde orbits. Photon circular radius defines the existence threshold limit $r>r_{ph}$ for timelike circular orbit which would of course be unstable. In the case of $a=0$, the above equations give $r_{ph}=\left(2\right)^{1/2}$ and $r_{ph}=\left(5/2\right)^{1/3}$ for prograde orbits in $D=5, 6$ respectively.

Since there occurs no stable orbits, the question of occurrence of ISCO does not arise. All marginally bound circular orbits given by $\mathcal{E}=1$ would be unbound, those with $\mathcal{E}<1$ falling into black hole while those with $\mathcal{E}>1$ escaping to infinity. When there exists the ISCO, energetically bound orbits with $\mathcal{E}<1$ on perturbation would climb to $r > r_{ISCO}$ and attain stable circular orbit.

\section{Pure Lovelock gravity}\label{sec:PLgravity}

{Lovelock gravity \cite{Lovelock71} is the most natural and fascinating generalization of Einstein gravity in higher dimensions. Lovelock action is a homogeneous polynomial of degree $N$ in Riemann curvature, which includes Hilbert-Einstein term in the linear order $N=1$, quadratic Gauss-Bonnet for $N=2$ and so on. The Lagrangian consists of dimensionally extended Euler densities and it is summed over all $N$, and each $N$ comes with a dimensionful coupling constant. Despite this it has the most remarkable property that on variation relative to metric it still yields the second order equation of motion. It is this feature that is unique to Lovelock generalization and that renders the theory most attractive as it is free of undesirable entities like ghosts. }

{Higher derivatives and higher dimensions are natural arena for string theory, and it turns out that one loop correction to the theory contains the Gauss-Bonnet term \cite{Zwiebach85,Sen05}. This gives a good foothold for GB/Lovelock as string inspired corrections to Einstein theory. The first vacuum solution of Einstein-Gauss-Bonnet equation was obtained in \cite{Boulware85} describing a static back hole. In general it is quite complicated to solve the equation for higher $N$, however static black hole solutions have been found \cite{Wheeler86a,Wheeler86b,Whitt88}.}

{Black string and black branes are extended structures covered by event horizon in the transverse direction and they could be easily constructed by adding Ricci flat directions to existing Einstein solutions. It turns out that this simple prescription does not work for Lovelock theory. It was then realized \cite{Kastor06,Giribet06} that for this to work one has to resort to pure Lovelock theory which involves only one $N$th order term without sum over the lower orders. On the other hand dimensionally continued black hole solutions were constructed where a relation between the coupling constants was prescribed, and in particular they were all given in terms of the unique vacuum $\Lambda$ \cite{BTZ1994,CTZ2000}. Pure Lovelock black hole solutions were obtained and studied for their thermodynamics in \cite{Cai06}.}

{This shows that there has been good motivation for pure Lovelock gravity which was further put on stronger formal ground in \cite{Dadhich12}. In here was defined Lovelock generalization of Riemann tensor \cite{Camanho16} with the property that trace of its Bianchi derivative vanishes giving a divergence free second rank symmetric tensor, an analogue of the Einstein tensor. This is exactly how Einstein tensor follows from the Riemann tensor. It is well known that Einstein gravity is kinematic in $D=3$, in the sense that Riemann curvature is entirely given in terms of Ricci, or equivalently, Weyl tensor is identically zero. This means there can exist no non-trivial vacuum solution in three dimension. Non-existence of non-trivial vacuum solution defines kinematicity of gravity. It is remarkable that this property could be universalized only in pure Lovelock gravity \cite{Dadhich12,Camanho16} to all critical odd $D=2N+1$ dimensions. That is in all $D=2N+1$, there can exist no non-trivial vacuum solution or equivalently Lovelock Riemann is given in terms of the corresponding Ricci and consequently Weyl is identically zero. Not only that bound orbits around a static black hole in higher dimensions could exist only in pure Lovelock theory and none else \cite{Dadhich13}. These two are the distinguishing characteristics of pure Lovelock gravity. It is worth noting that vacuum Einstein gravity in $D=4$ dimensions is similar to pure Lovelock theory in D = 3N +1 dimensions \cite{Chakraborty18,Gannouji19}. Following \cite{Dadhich14,Dadhich16a}, the compactness limit for stable stellar structures in pure Lovelock theories was studied
in $D \geq 2N + 1$ dimensions \cite{Chakraborty20egb,Dadhich17,Dadhich17b}. The Brown-York quasi-local energy was studied for pure Lovelock black holes in~ \cite{Chakraborty15} while the role of gravitational field energy was brought about in defining the Buchdahl compactness bound for static fluid spheres in~\cite{Dadhich20:JCAP}. }

{Besides black string and black brane solutions, universality of kinematic property for all critical odd $D=2N+1$ dimensions and existence of bound orbits around static black hole uniquely pick out pure Lovelock theory. Thus the theory is well grounded and strongly motivated as the proper gravitational theory in higher dimensions \cite{Dadhich16a}.}

\section{Pure GB rotating black hole and orbits 
}\label{sec:GB}

Unlike the Myers-Perry solution, there exists no exact solution of pure GB vacuum equation describing a rotating black hole. However a pure GB rotating black hole metric was constructed in~\cite{Dadhich-Ghosh13}. It was based on the novel and innovative method \cite{Dadhich13b} by which the Kerr metric was obtained without solving the field equations. It appeals to the two physically motivated properties, a photon falling along the axis of rotation experiences no three acceleration while a timelike particle experiences the Newtonian acceleration. Of course for this one has to choose an appropriate $3$-geometry, spherical for non-rotating and ellipsoidal for rotating black hole. When this method was applied to an appropriate ellipsoidal metric in six dimension, and on replacing Newtonian acceleration in the above construction by pure GB acceleration, the metric that followed had all the required properties of a rotating black hole. Unfortunately it failed to satisfy the pure GB vacuum equation. It does however do satisfy the equation in the leading order.

{It thus turns out that inclusion of Newtonian potential into an appropriate rotating spacetime leads to the Kerr metric of rotating black hole. Recently this method has been successfully employed  \cite{Amin-Mirza21} to obtain the Myers-Perry metric for rotating black hole in higher dimensions in which higher dimensional potential $\mu/r^{D-3}$ replaces $\mu/r$ in appropriate axially symmetric rotating spacetime. Unfortunately this straight forward and physically appealing method does not quite work for pure Lovelock gravity. This may be due to equation being non-linear in Riemann curvature. It is nevertheless physically well motivated where pure Lovelock potential is incorporated into the corresponding ellipsoidal rotating geometry. It should therefore represent a valid rotating black hole metric which may not though be an exact solution of pure Lovelock equation. Yet it should be taken as a good rotating black hole spacetime that incorporates pure Lovelock potential. Its energetics and optical properties have also been studied in \cite{Abdujabbarov15a}. Recently the metric for $4D$-Einstein-GB rotating black hole has similarly been obtained~\cite{Kumar20egb} and it is not an exact solution of the equation. We shall take this metric to study six dimensional pure GB rotating black hole not in full at all orders in the next section. }

{One of the distinguishing properties of pure Lovelock is that gravity is kinematic in all critical odd $D=2N+1$ dimensions; i.e. Lovelock analogue of Riemann curvature is entirely given in terms of the corresponding Ricci~\cite{Dadhich12,Camanho16} so that vacuum solution is trivial. Therefore for non-trivial vacuum solution, dimension of spacetime has always to be $>2N+1$.}

The pure GB six dimensional rotating black hole~\cite{Dadhich-Ghosh13} is described by the metric
\begin{eqnarray}\label{Eq:GB_metric}
ds^2&=&-\frac{\Delta}{\Sigma}\left(dt-a_1\sin^2\theta
d\phi-a_2\cos^2\theta
d\psi\right)^2+\frac{\Sigma}{\Delta}dr^{2}\nonumber\\
&+&\Sigma d\theta^2 +
\frac{\sin^2\theta}{\Sigma}\left[(r^2+a_1^2)d\phi-a_1
dt\right]^2\nonumber\\&+&\frac{\cos^2\theta}{\Sigma}\left[(r^2+a_2^2)d\psi-a_2
dt\right]^2 \nonumber\\&+&
r^2\left(\cos^2\theta+\sin^2\phi\right)d\Omega^2_{D-4}\ ,\ \quad
\end{eqnarray}
with
\begin{eqnarray}\label{Eq:D11}
 \Sigma &=&r^2+a_1^2\cos^2\theta+a_2^2\sin^2\theta\, , \nonumber\\
\Delta &=& \frac{(r^2+a_1^2)(r^2+a_2^2)}{r^2} -2\mu r^{2-\alpha} \, ,
\end{eqnarray}
where $\alpha = (D-2N-1)/N = 1/2$ for $D=6$.

Black hole horizons would be given by $\Delta = 0$
which for single rotation reduces to
\begin{eqnarray}
r^4 - 4\mu^2 r^3 + 2a^2r^2 + a^4 = 0\, .
\end{eqnarray}
It solves to give
 \begin{eqnarray}\label{lovelock_hor1}
 r_{\pm}&=&{\mu^2}+ \frac{X({\mu},a)}{\sqrt{6}}\pm
 \left({8\mu^4}-\frac{8a^4-2Y^2({\mu},a)}{3Y(\mu,a)}\right.\nonumber\\&+&\left. \sqrt{3}\frac{64\mu^6-32 a^2 \mu^2)}{4\sqrt{2}X(\mu,a)}-\frac{8a^2}{3}\right)^{1/2}\, ,
 \end{eqnarray}
with
 \begin{eqnarray}
 X^{2}(\mu,a)&=& 6\mu^4-2a^2+ \frac{4a^{4}}{Y(\mu,a)}+Y(\mu,a)\, ,
 \nonumber\\
Y^{3}(\mu,a)&=& {27}\mu^4 a^4 -8a^6+3\sqrt{3}\,\mu^2\,
a^4\sqrt{{27}\mu^4-16 a^{2}}\nonumber \, .
 \end{eqnarray}

As said earlier, this is not a solution of the pure GB vacuum equation yet it has all the desired properties and the equation is though satisfied in the leading order. It however describes a rotating black hole.

In Einstein gravity which is linear order $N=1$ pure Lovelock, bound orbits around a static black hole can exist only in four dimension and none else~\cite{Dadhich13}; i.e. no bound and thereby no stable circular orbits can occur in all higher dimensions. Such orbits in higher dimensions are only provided by pure Lovelock gravity in dimensions, $2N+2 \leq D \leq 4N$; i.e. for pure GB in $D= 6, 7, 8$. The lower limit follows from the requirement of gravity being dynamic while the upper one from $\alpha < 2$ so that centrifugal force is able to balance gravitational attraction to give bound orbits.

{With $\Delta$ written for pure GB rotating black hole with single rotation in the above metric (\ref{Eq:GB_metric}), the effective potential
is given by
\begin{eqnarray}
V_{eff}&=& \frac{\Phi}{\Psi} \frac{a \mathcal{L}}{r^2}
+ \frac{\sqrt{\Big(\Psi + \frac{\mathcal{L}^2}{r^2}\Big)\Big(1 + \frac{a^2}{r^2} - \Phi\Big)}}{\Psi} \, ,
\end{eqnarray}
where
\begin{eqnarray}
\Phi = \frac{M}{r^{\big(D-2N-1\big)/N}}\, \mbox{~and~} \Psi = 1 + \frac{a^2}{r^2} + \Phi \frac{a^2}{r^2} \, .
\end{eqnarray}
The effective potential in the above would in particular for $D=6,8,9$ dimensions reads as}
\begin{eqnarray}\label{Eq:Veff_GB}
V_{eff;6D}^{GB}(r)&=&\frac{2 a \mu \mathcal{L}}{r^{5/2}+\left(r^{1/2}+2\mu\right)a^2}\nonumber\\&+&\frac{\Big(r^3+\left(r+2\mu r^{1/2}\right)a^2 +r \mathcal{L}^2\Big)^{1/2}}{r^{5/2}+\left(r^{1/2}+2\mu\right)a^2}\nonumber\\&\times &\Big(r\left(r-2\mu r^{1/2}\right)+a^2 \Big)^{1/2}\, ,
\end{eqnarray}
%
\begin{figure*}
\centering
 \includegraphics[width=0.45\textwidth]{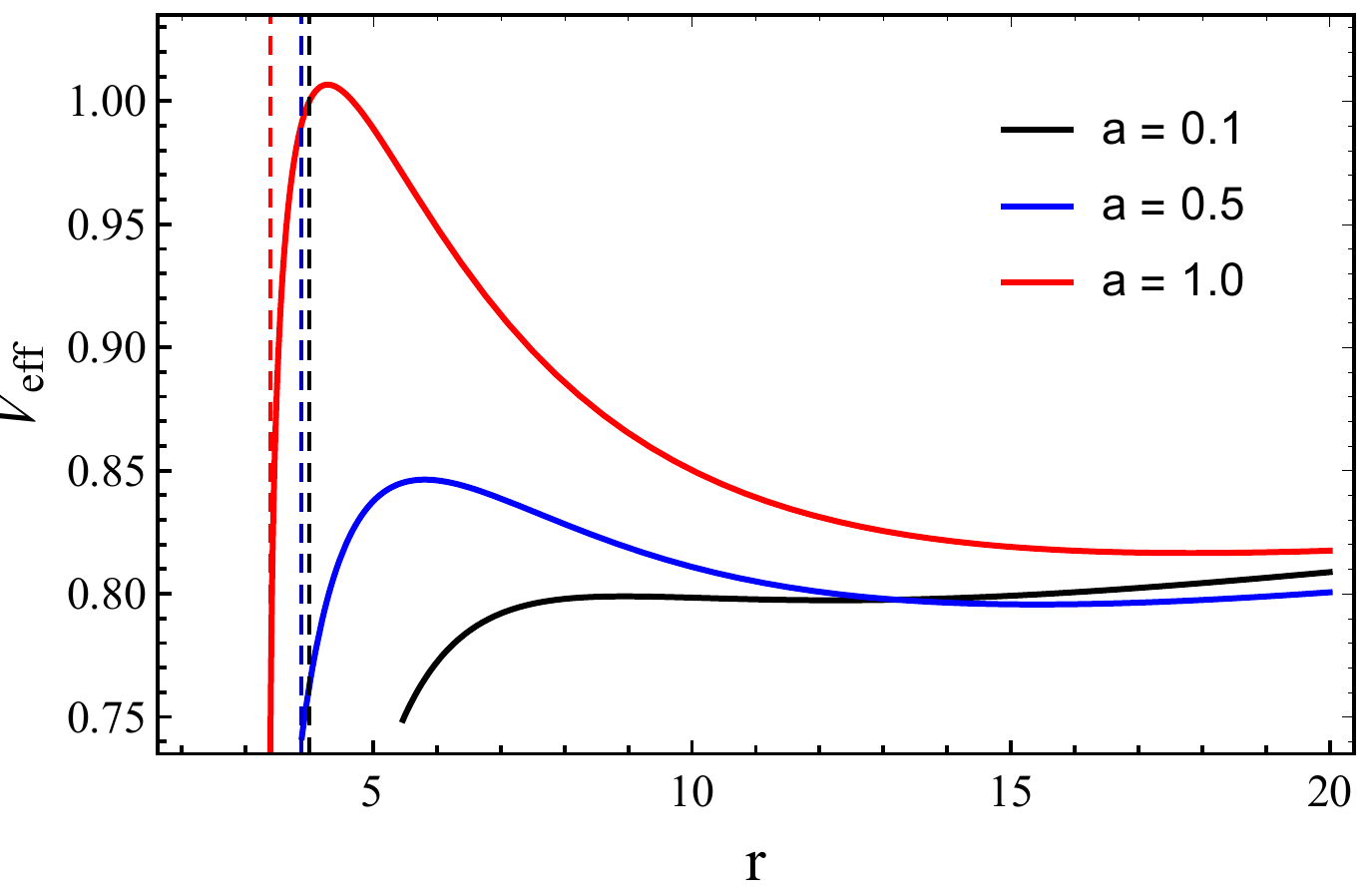}
 \includegraphics[width=0.45\textwidth]{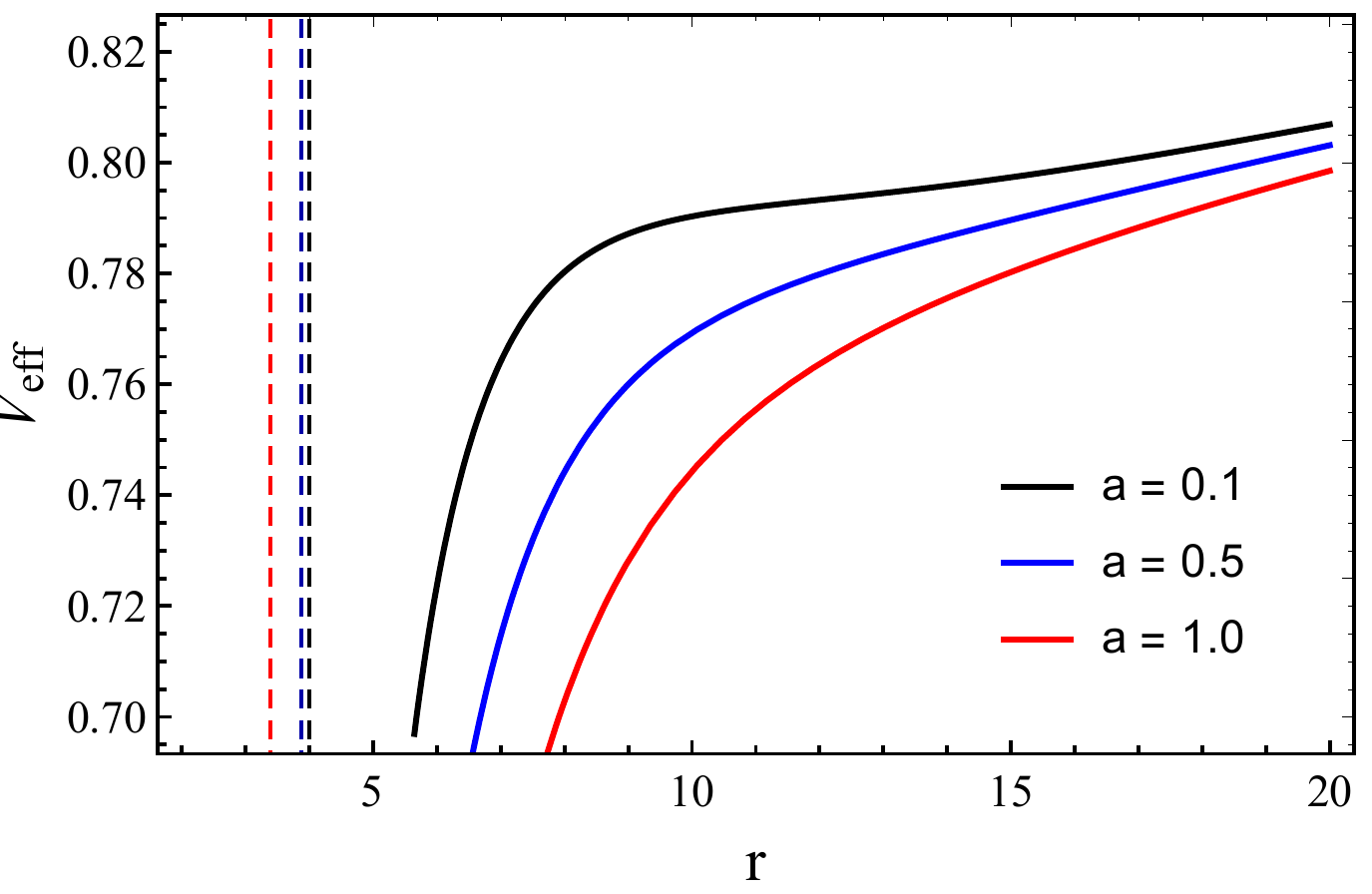}

 \caption{\label{fig:GB} {$V_{eff}$ for $D=6$, prograde (left) and retrograde (right) orbits for $\mathcal{L}= \pm 8.5$. Vertical lines indicate location of horizon.}}
\end{figure*}
\begin{figure*}
\centering
 \includegraphics[width=0.45\textwidth]{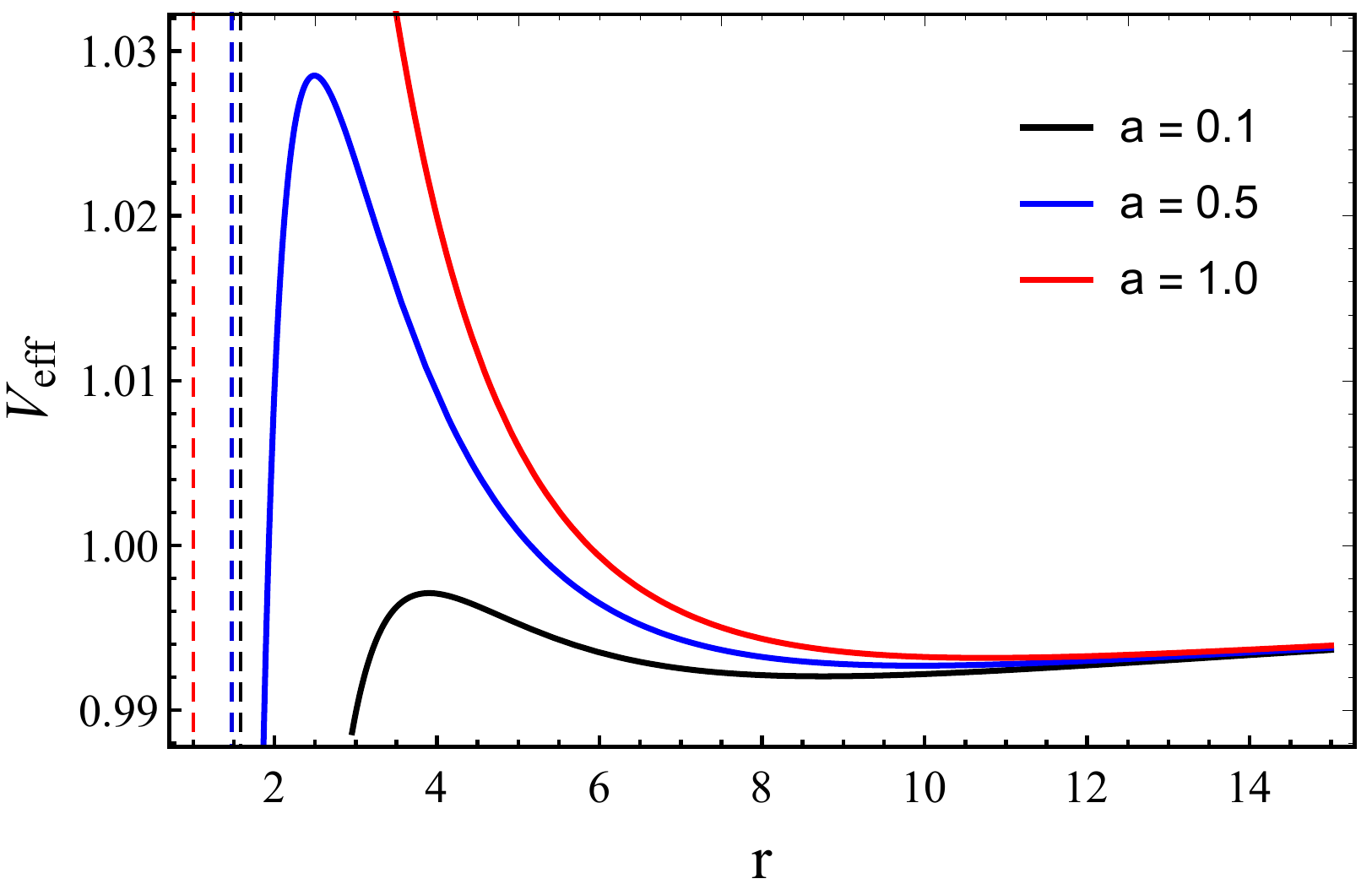}
 \includegraphics[width=0.45\textwidth]{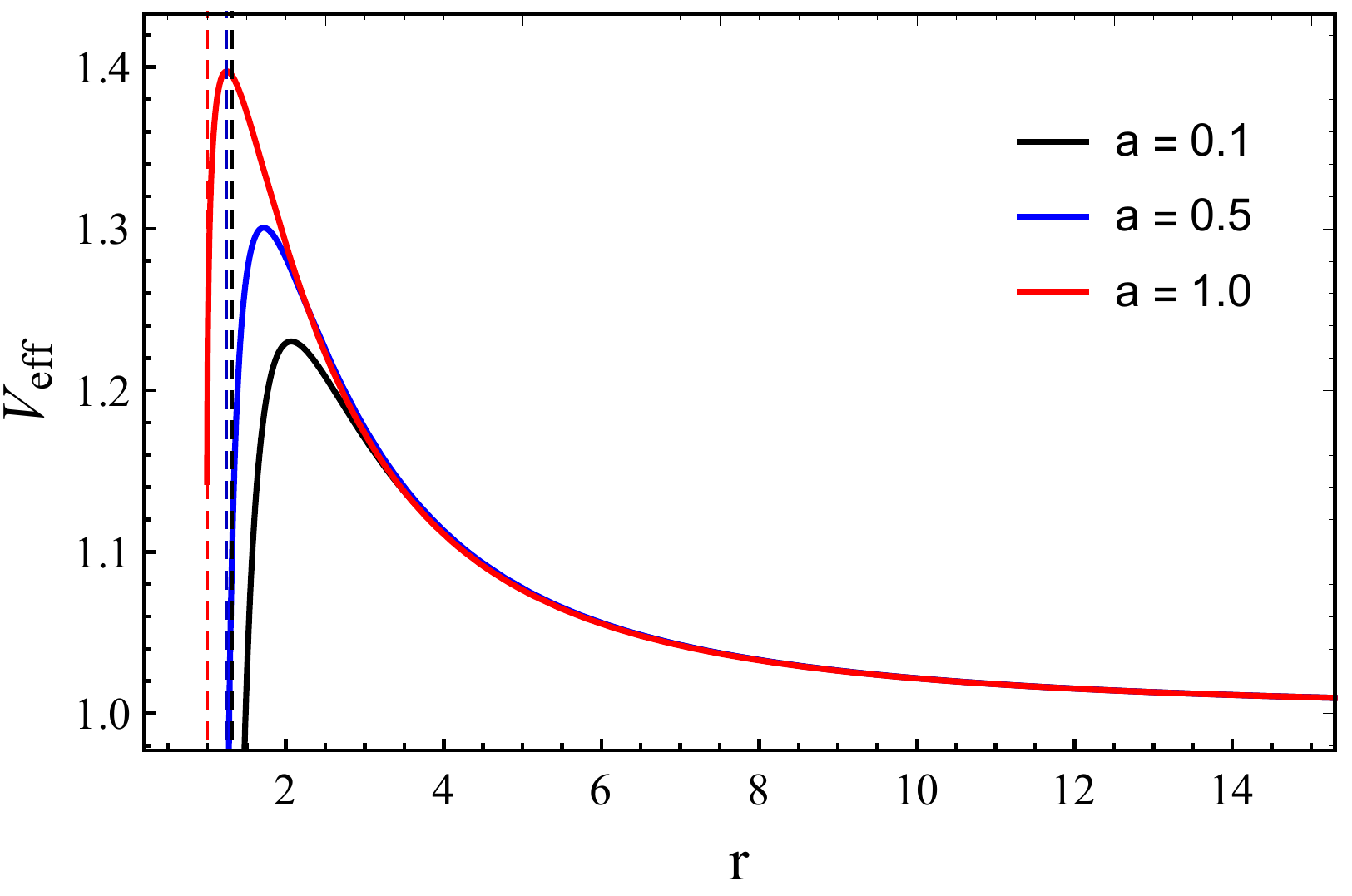}

 \caption{\label{fig:GB2} Left and right panels show $V_{eff}$ for $\mathcal{L}=2.25$ in $D= 8, 9$ respectively. Vertical lines indicate location of horizon. }
\end{figure*}
\begin{figure*}
\centering
 \includegraphics[width=0.45\textwidth]{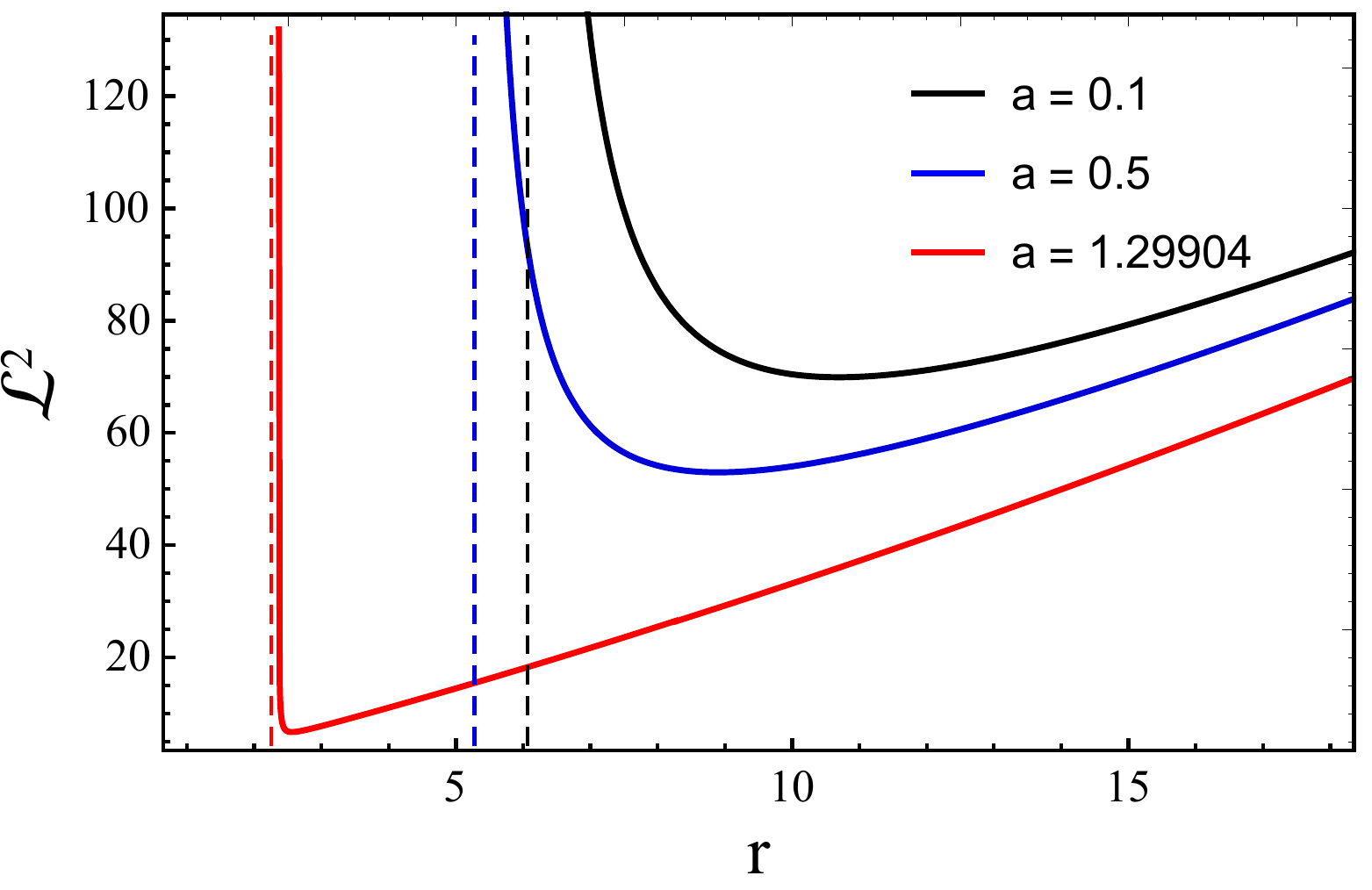}
 \includegraphics[width=0.45\textwidth]{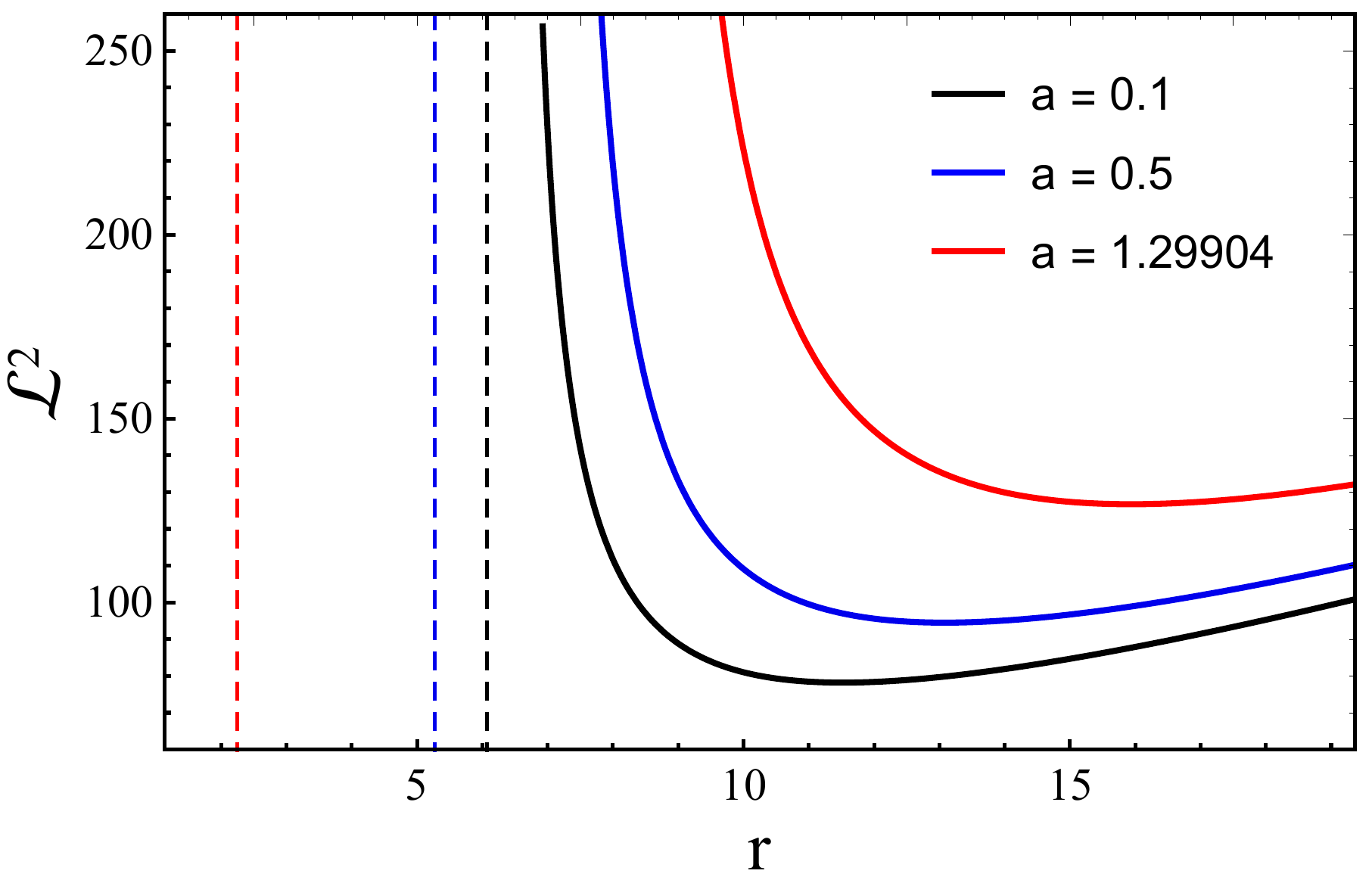}

 \includegraphics[width=0.45\textwidth]{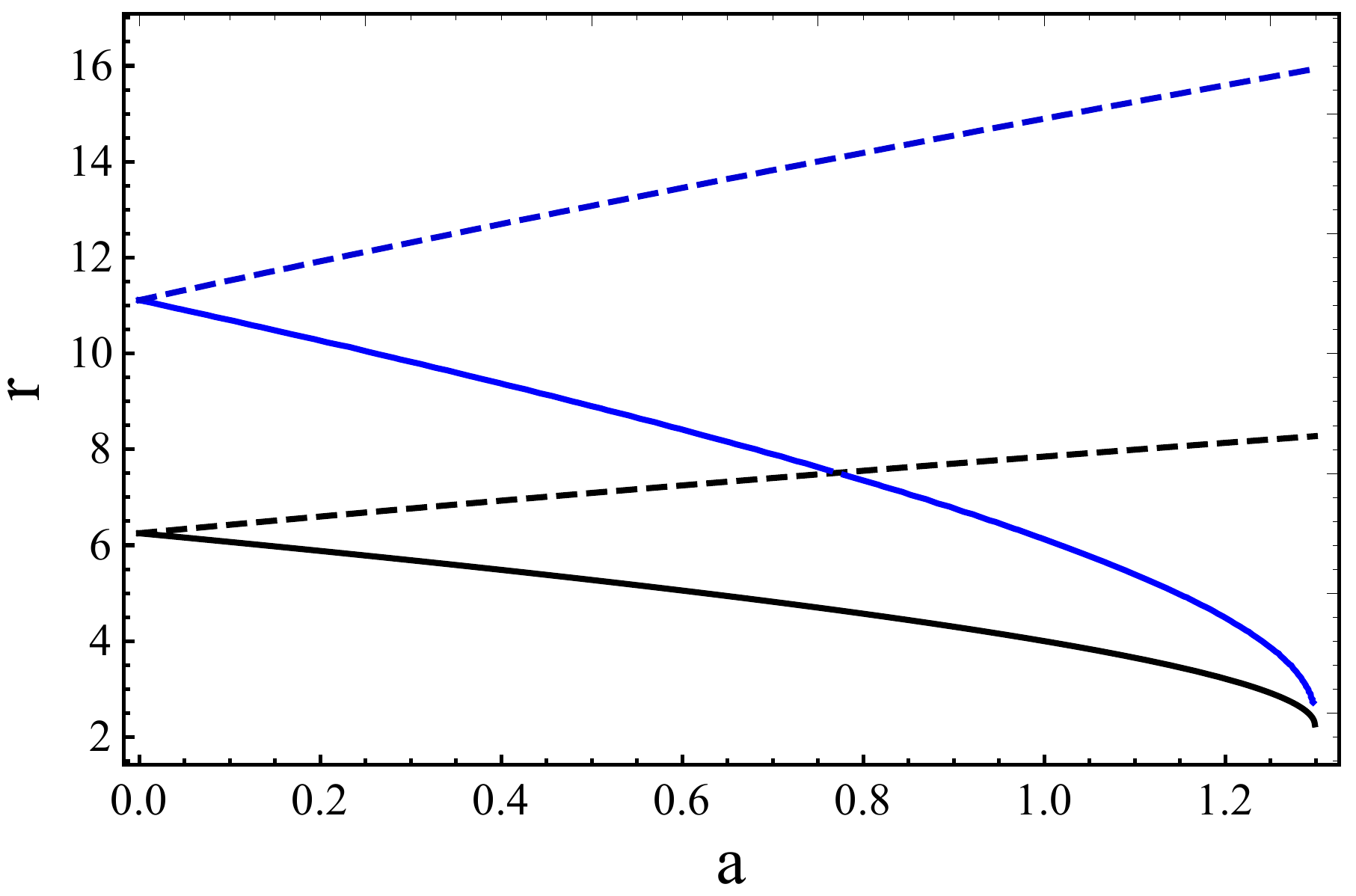}

 \caption{\label{fig:GBL} {Top row, left panel: $\mathcal{L}^2$ profile is plotted for prograde orbits. Top row, right panel: $\mathcal{L}^2$ profile is plotted for retrograde orbits. Vertical lines in the top row show location of $r_{ph}$. Bottom panel: $r_{ph}$ (black line) and ISCO radius (blue line) plotted against rotation parameter $a$, thick and dashed lines respectively show prograde and retrograde orbits. Note that $r_{ph}=6.25$ for $a=0$.}}
\end{figure*}
and
%
\begin{eqnarray}
V_{eff;8D}^{GB}(r)&=&\frac{2 a \mu \mathcal{L}}{r^{7/2}+\left(r^{3/2}+2\mu\right)a^2 }\nonumber\\&+&\frac{ \Big(r^5+\left(r^3+2\mu r^{3/2}\right)a^2 +r^3 \mathcal{L}^2\Big)^{1/2}}{r^{7/2}+\left(r^{3/2}+2\mu\right)a^2 }\nonumber\\&\times &\Big(r^2-2\mu r^{1/2}+a^2 \Big)^{1/2}\, ,\\
V_{eff;9D}^{GB}(r)&=& \frac{2 a \mu \mathcal{L}}{r^{4}+\left(r^{2}+2\mu\right)a^2 }\nonumber\\&+&\frac{ \Big(r^6+\left(r^4+2\mu r^{2}\right)a^2 +r^4 \mathcal{L}^2\Big)^{1/2}}{r^{4}+\left(r^{2}+2\mu\right)a^2 }\nonumber\\&\times &\Big(r^2-2\mu +a^2 \Big)^{1/2}\, ,
\end{eqnarray}
respectively for $D = 8, 9$.
In the limit of large $r$, these expressions reduce to
\begin{eqnarray}
V_{eff;6D}^{GB}(r\to r_{\infty})&\sim & 1-\frac{\mu}{r^{1/2}}- \frac{\mu^2}{2r}-\frac{\mu^3}{r^{3/2}}\nonumber\\&+&\left(\frac{\mathcal{L}^2}{2r^2}-\frac{5 \mu^4}{8r^2}\right)+\mathcal O\left(\frac{1}{r^{5/2}}\right)\, ,\\
V_{eff;8D}^{GB}(r\to r_{\infty})&\sim & 1-\frac{\mu}{ r^{3/2}} -\frac{\mu^2 }{2r^3}+\frac{\mathcal{L}^2}{2r^2 }+\mathcal O\left(\frac{1}{r^{7/2}}\right)\, ,\nonumber\\ \\
V_{eff;9D}^{GB}(r\to r_{\infty})&\sim & 1+\frac{\mathcal{L}^2}{2r^2 } -\frac{\mu}{r^{2}} -\frac{a^2 \mathcal{L}^2}{2r^4}+\mathcal O\left(\frac{1}{r^{9/2}}\right)
\, , \nonumber\\
\end{eqnarray}

Note that for the former two (as well as for $D=7$ for which potential falls off as $1/r$), $V_{eff}$ tends to unity at infinity from the below while for the latter from the above. This means bound orbits can exist only in $D = 6, 7, 8$. For bound orbits to exist there must occur a potential well with a minimum, $V_{eff} < 1$ there. This cannot happen for $D> 4N$ in general and $D> 8$ in particular for pure GB.

Let us first consider the particular case of $D=6$. For circular orbits, we should have
\begin{eqnarray}\label{Eq:first_der1}
 \frac{\partial V_{eff}(r)}{\partial r}=0\, ,
\end{eqnarray}
giving the equation,
\begin{eqnarray}\label{Eq:GB1}
&&\frac{\left(r^{5/2}+a^2 \left(2 \mu +\sqrt{r}\right)\right)\left(r^2-2 \mu  r^{3/2}+a^2\right)^{-1/2} }{ \big(r^3+r \mathcal{L}^2+a^2 \left(2 \mu  \sqrt{r}+r\right)\big)^{1/2}}\nonumber\\ &\times &\Bigg[a^4 \left(\mu +\sqrt{r}\right)+a^2 \sqrt{r} \left(6 r^2-8 \mu ^2 r+\mathcal{L}^2\right)+3 r^{5/2} \mathcal{L}^2\nonumber\\&+&5 r^{9/2}-9 \mu  r^4-5 \mu  r^2 \mathcal{L}^2\Bigg]\nonumber\\&-&\left(a^2+5 r^2\right) \left[2 a \mu  \mathcal{L}+\Big(a^2-2 \mu  r^{3/2}+r^2\Big)^{1/2}\right.\nonumber\\&\times & \left. \Big(r^3+r \mathcal{L}^2+a^2 \left(2 \mu  \sqrt{r}+r\right)\Big)^{1/2}\right]=0 \, .
\end{eqnarray}
This equation has two positive roots for which $V_{eff}$ attains maximum and minimum (see Fig.~\ref{fig:GB}, left panel). This would be so for $D = 7, 8$ as well because the allowed range for bound orbits for pure GB is $6 \leq D \leq 8$. In particular for $D=7$ (in general for $D=3N+1$ \cite{Chakraborty18}), potential has the same fall off, $1/r$ as for the Kerr black hole, and hence the effective potential would be the same as for the Kerr metric in four dimension. However for $D\geq 9$, there would occur only a maximum indicating absence of bound orbits and consequently also of stable circular orbits (see~Fig.~\ref{fig:GB2}, right panel).

\begin{figure*}
\centering
\includegraphics[width=0.45\textwidth]{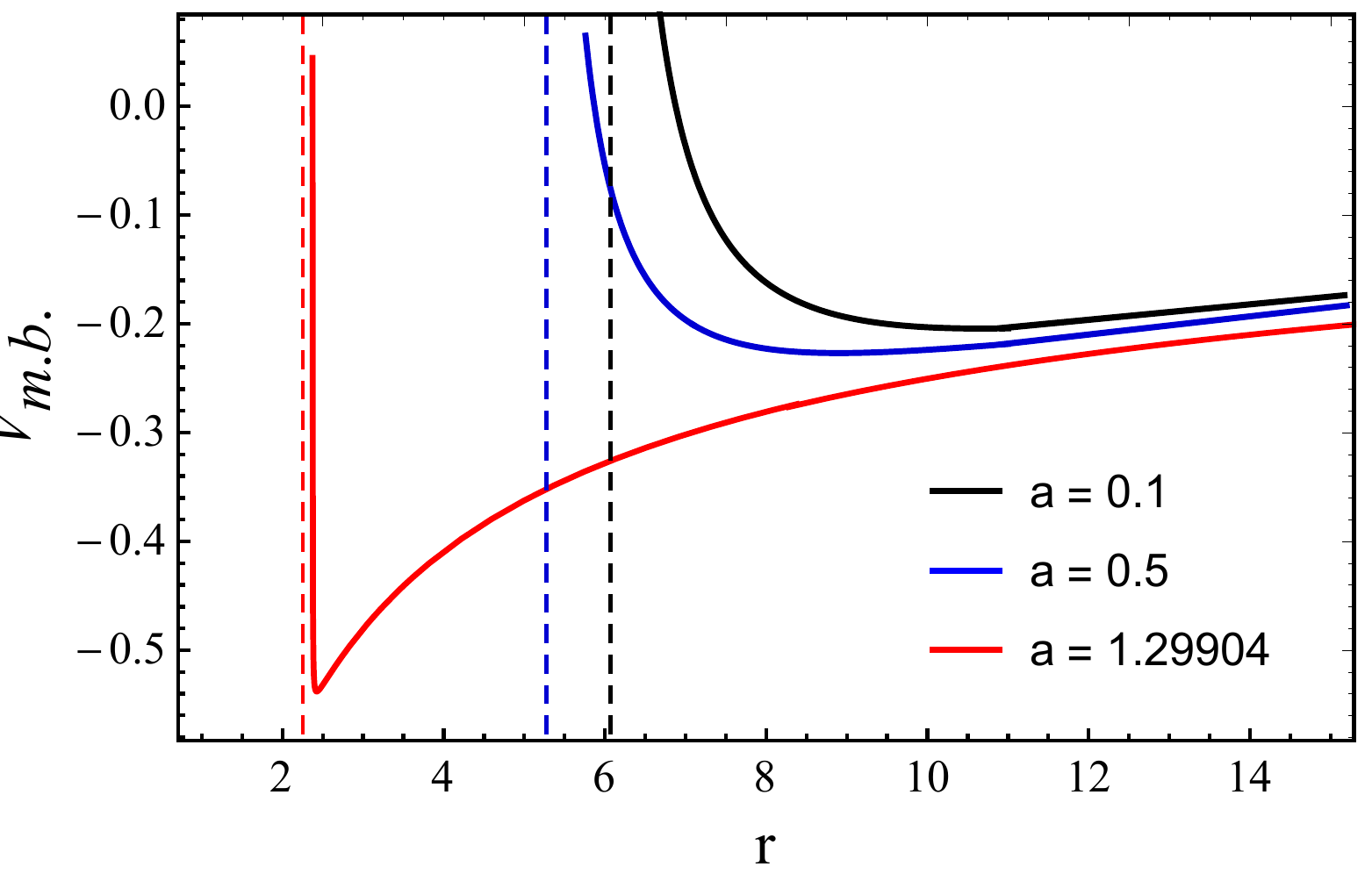}
 \includegraphics[width=0.45\textwidth]{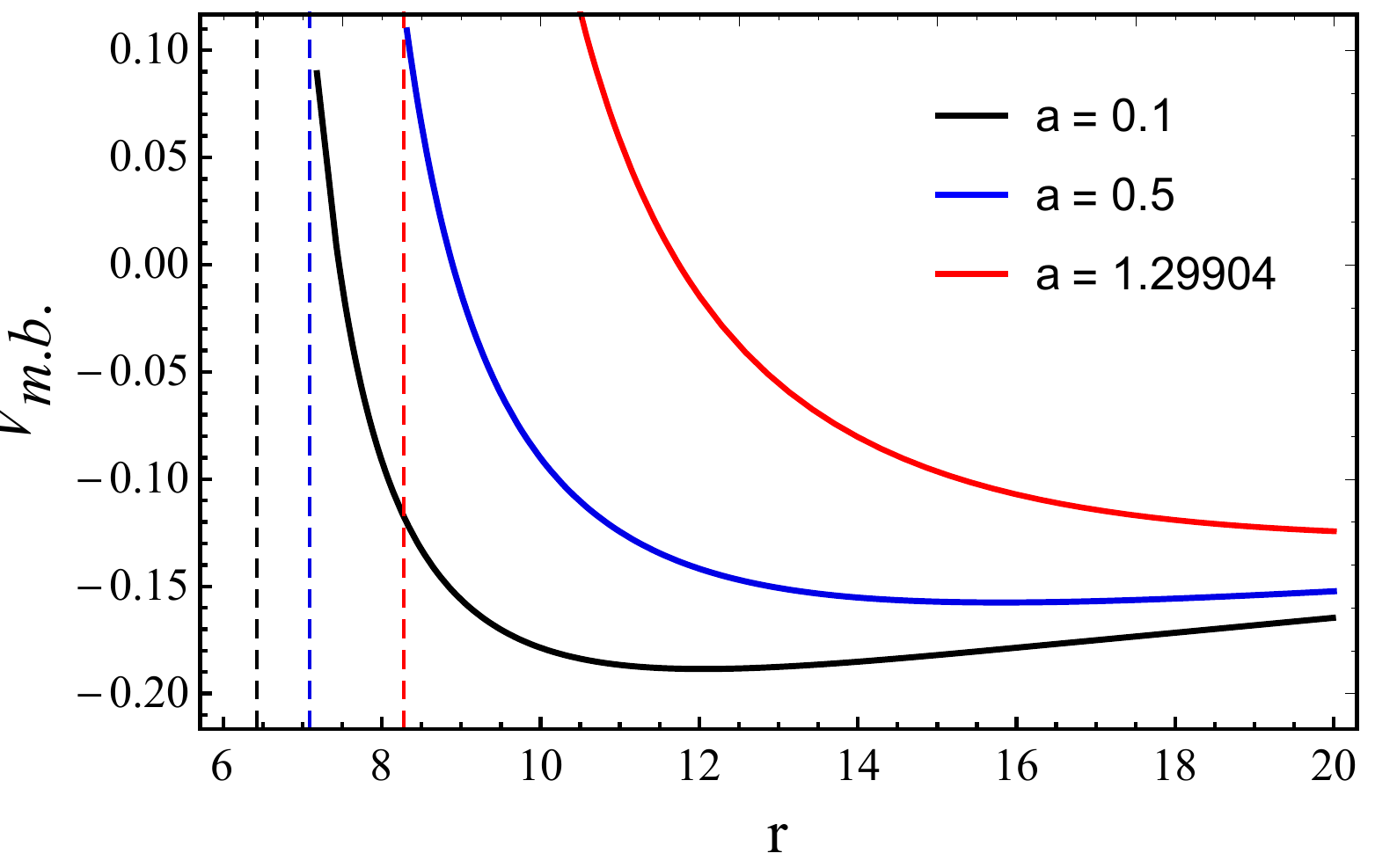}

 \caption{\label{fig:mb1} {Marginally bound circular orbits in $D=6$: Left and right panels respectively refer to prograde and retrograde orbits. Vertical lines indicate location of $r_{ph}$. }}
\end{figure*}
Solving the above equation~(\ref{Eq:GB1}) for the angular momentum we write,
\begin{eqnarray}\label{Eq:An3}
\mathcal{L}^2_{\pm\,GB}(r)=\frac{\mu r FG \mp 2 \sqrt{2} \left(a^2-2 \mu  r^{3/2}+r^2\right)N}{r^{3/2} \left(r^{3/2} \left(2 \sqrt{r}-5 \mu \right)^2-8 a^2 \mu \right) F}\, ,
\end{eqnarray}
where
\begin{eqnarray}\label{Eq:fgn}
F&=&\left(r^{5/2}+a^2 \left(\sqrt{r}+2 \mu \right)\right)^2 \, ,\nonumber\\
G&=&2 \sqrt{r} \left(r^2+a^2\right)^2+\mu  \left(22 a^2 r^2+11 a^4-5 r^4\right)\nonumber\\&-&40 \mu ^2 a^2 r^{3/2}\, ,\nonumber\\
N&=&\left(\mu ^3 \sqrt{r} \left(a^3+5 \,a r^2\right)^2\,F^2 \right)^{1/2}\, .
\end{eqnarray}

Fig.~\ref{fig:GBL} (top row) shows its radial profile where minimum indicates the ISCO radius. This defines the lower bound on angular momentum for giving rise to stable circular orbits; i.e. particles with angular momentum lower than this threshold value would fall in carrying angular momentum to black hole. Therefore existence of this bound on angular momentum; i.e. ISCO is of primary critical importance for formation of rotating black holes \cite{Dadhich20}. ISCO radius however shifts toward the horizon with increasing value of $a$.

Further we explore the threshold value of angular momentum numerically, which is defined by the ISCO; see Table~\ref{tab1}. As expected the ISCO radius and corresponding angular momentum decrease with increasing rotation parameter $a$.

\begin{table}
{\caption{\label{tab1} $r_{min} (r_{ISCO})$ and $\mathcal{L}_{min}$ are tabulated for different values of rotation $a$ for prograde and retrograde rotating orbits. Note that $a_{extremal} = 1.29904$.}
\begin{tabular}{c|cccc}
$\rm a $  & $r_{min}(pro)$
& $\mathcal{L}_{min}$ & $r_{min}(retro)$ & $\mathcal{L}_{min}$  
\\ \hline \\
%
%
%
0.0    &11.1111 &8.60663 &11.1111  &8.60663    \\\\
0.1    &10.6931 &8.36242 &11.5197  &8.84274     \\\\
0.5    &8.89855 &7.27935 &13.0779  &9.72136   \\\\
1.0    &6.12307 &5.46058 &14.8970  &10.7091    \\\\
1.29904 & 2.2500 &2.32379 &15.9345 &11.2565     \\
\end{tabular}}
\end{table}

For pure GB rotating black hole, ISCOs will only occur in $D= 6, 7,8$ and not in any dimension $> 8$.

Let us next consider another limit on existence of the circular geodesics. The ISCO defines the stability threshold and below which would occur unstable orbits which would naturally be bounded from below by the photon circular orbit. That is defined by $\mathcal{L}^2_{\pm}\rightarrow\infty$, and that gives
\begin{eqnarray}\label{Eq:ph_con}
r^{3/2} \left(2 \sqrt{r}-5 \mu \right)^2-8 a^2 \mu =0\, .
\end{eqnarray}
For $a=0$, $r_{ph}/\mu^2 = 25/4$ while for $a\neq 0$ its location is shown in Fig.~\ref{fig:GBL} (bottom row).

Note that the existence threshold for circular orbit is therefore given by $r > r_{ph}$ while the stability threshold is given by $r \geq r_{ISCO}$. In between these two lies the marginally (energetically) bound threshold which is given by
\begin{eqnarray}\label{Eq:mb}
\mathcal{E}(r_{mb})=1\, ,
\end{eqnarray}
where $\mathcal{E}=E/m$ refers to energy of particle per unit mass. Orbits are energetically bound for $\mathcal{E} \leq 1$ and are unbound for $\mathcal{E} > 1$.

All orbits occurring between $r_{ph} < r < r_{ISCO}$ are unstable, and their ultimate fate is decided by the marginally bound condition, $V_{eff} = 1$. Let us define $V_{eff}(mb) = V_{eff} - 1$, when it is positive, on perturbation particle escapes to infinity and when it is negative, it climbs up to $r >r_{ISCO}$ to a stable circular orbit. The radius threshold for marginally bound orbit, $r_{mb}$ is given by the solution of $V_{eff}(mb) = 0$. So for $r_{ph} < r < r_{mb}$ orbit is unbounded while for $r >r_{mb}$ it is bounded. We explore the radius threshold for marginally bound orbit numerically in Table~\ref{tab2} for prograde and retrograde rotating orbits.

For $D=6$, on substituting Eq.~(\ref{Eq:An3}) in Eq.~(\ref{Eq:Veff_GB}) we obtain
\begin{eqnarray}\label{Eq:mb_gb}
V^{\pm{GB}}_{mb}(r)&=&\frac{2 a \mu }{r^{5/2}+\left(r^{1/2}+2\mu\right)a^2 }\nonumber\\&\times &\left(\frac{\mu r FG \mp 2 \sqrt{2} \left(a^2-2 \mu  r^{3/2}+r^2\right)N}{r^{3/2} \left(r^{3/2} \left(2 \sqrt{r}-5 \mu \right)^2-8 a^2 \mu \right) F}\right)^{1/2}\nonumber\\&-& 1 +\frac{\Big(r\left(r-2r^{1/2}\mu\right)+a^2 \Big)^{1/2} }{r^{5/2}+\left(r^{1/2}+2\mu\right)a^2 }\nonumber\\&\times &\Bigg[r^3+\left(r+2\mu r^{1/2}\right)a^2 \nonumber\\&+& \frac{\mu r FG \mp 2 \sqrt{2} \left(a^2-2 \mu  r^{3/2}+r^2\right)N}{r^{1/2} \left(r^{3/2} \left(2 \sqrt{r}-5 \mu \right)^2-8 a^2 \mu \right) F}\Bigg]^{1/2}\, ,
\end{eqnarray}
where $F$, $G$ and $N$ are given in Eq.~(\ref{Eq:fgn}).
We then plot the radial profile of $V^{\pm}_{mb}$ for pro- and retrograde-rotating orbits in Fig.~\ref{fig:mb1}.

\begin{table}
\caption{\label{tab2} {The values of $r_{mb}$  are tabulated for different values of rotation $a$ for prograde and retrograde rotating orbits.}}  
\begin{tabular}{c|cc}
$\rm a $  & ${r_{mb}(prograde)}$
& ${r_{mb}(retrograde)}$  
\\ \hline \\
0.0    &7.11111  &7.11111    \\\\
0.1    &6.87725  &7.46032     \\\\
0.5    &5.86838 &8.89124   \\\\
1.0    &4.29261 &10.6881    \\\\
1.29904  & - &11.7449     \\
\end{tabular}
\end{table}

From Fig.~\ref{fig:mb1}, one can see that for $r >r_{mb}$, $V^{\pm}_{mb}<0$ for pro-and retrograde rotating orbits, thereby indicating that orbits remain bounded on perturbation. The opposite is the case for $r < r_{mb}$.
\begin{figure*}
	\centering
	\includegraphics[width=0.45\textwidth]{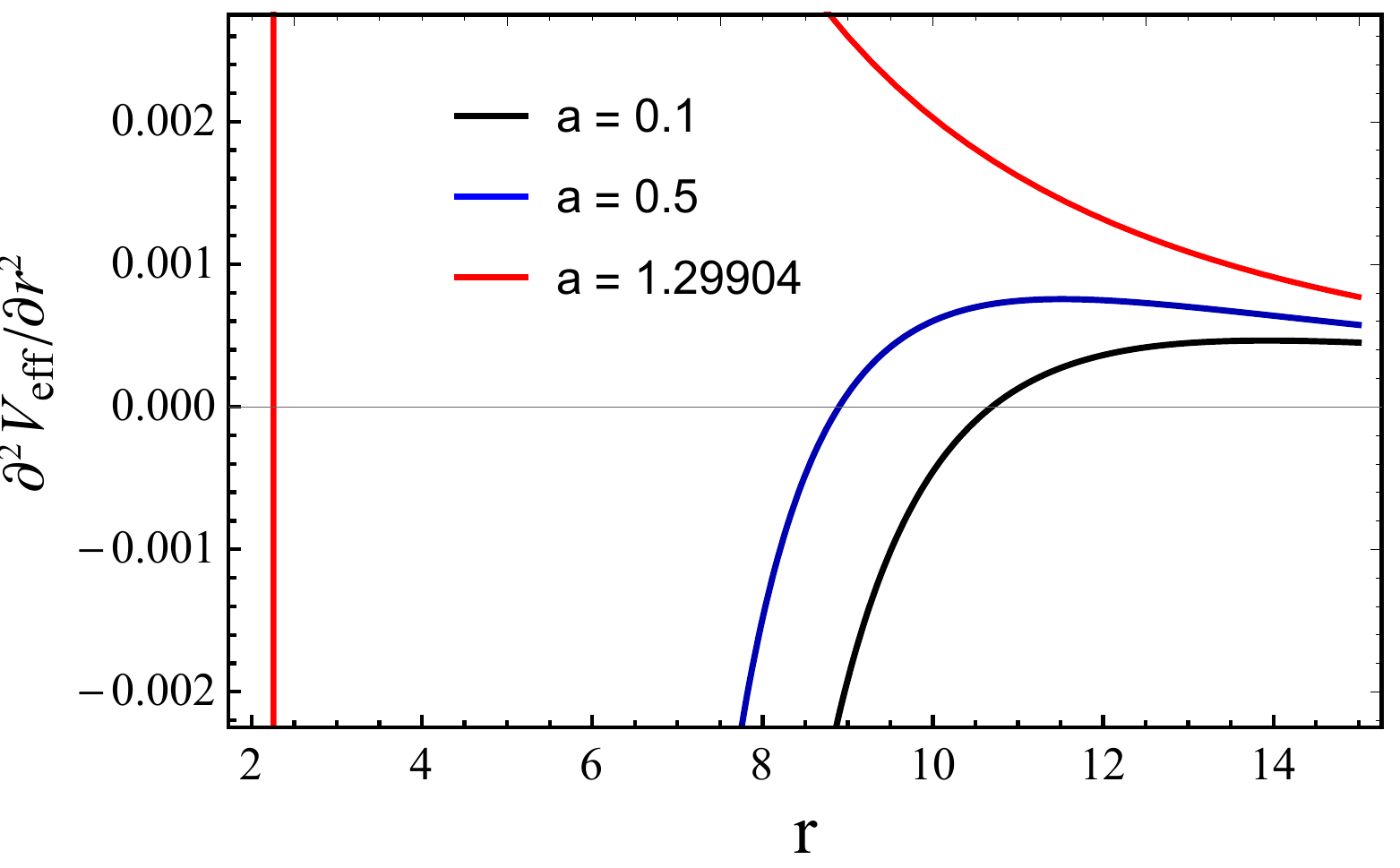}
	\includegraphics[width=0.45\textwidth]{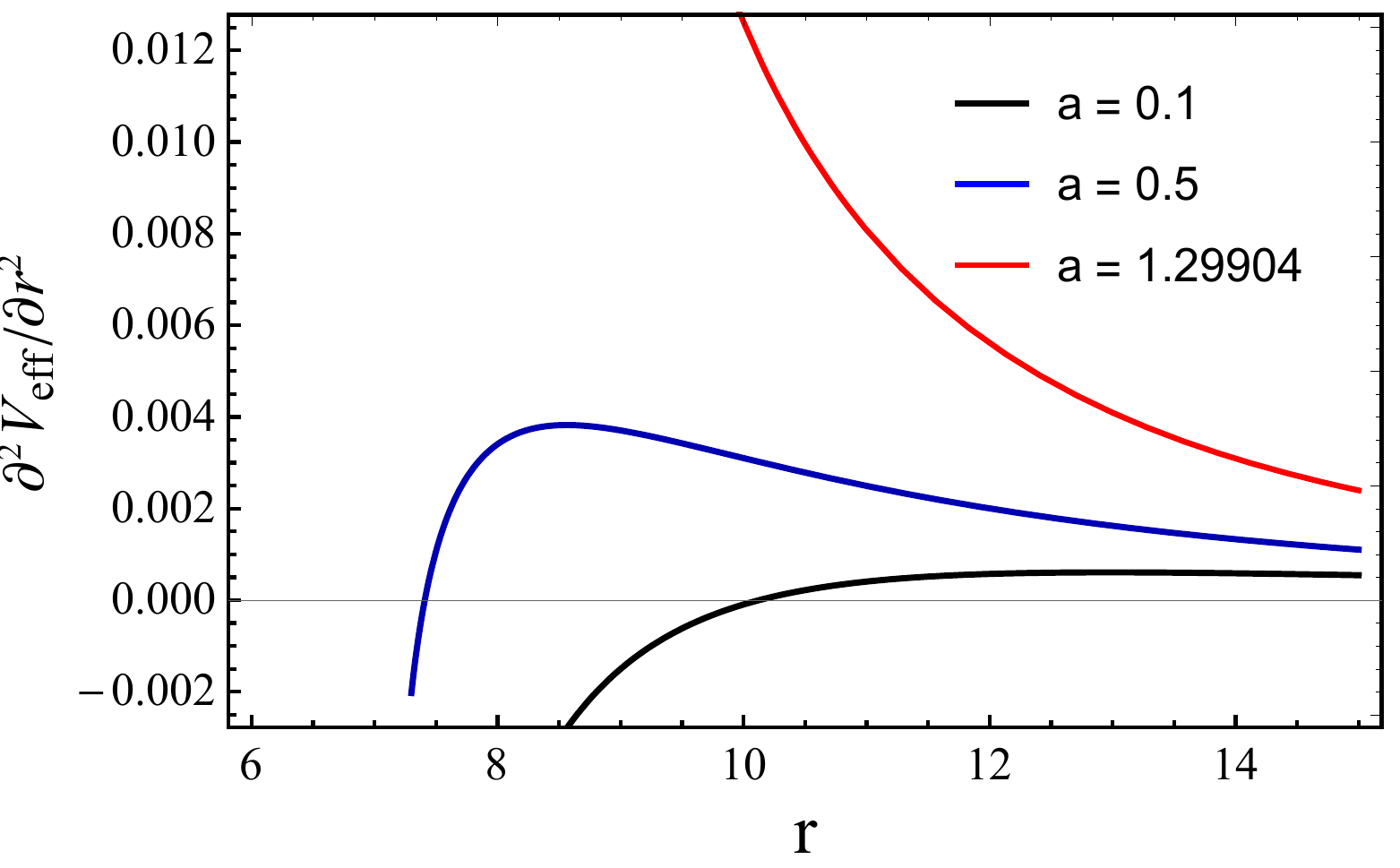}
	
	\caption{\label{fig:dif} {$V_{eff}^{\prime\prime}$ is plotted  where Left/right panels refer to prograde/retrograde rotating orbits.  }}  \end{figure*}
\begin{figure*}
\centering
 \includegraphics[width=0.45\textwidth]{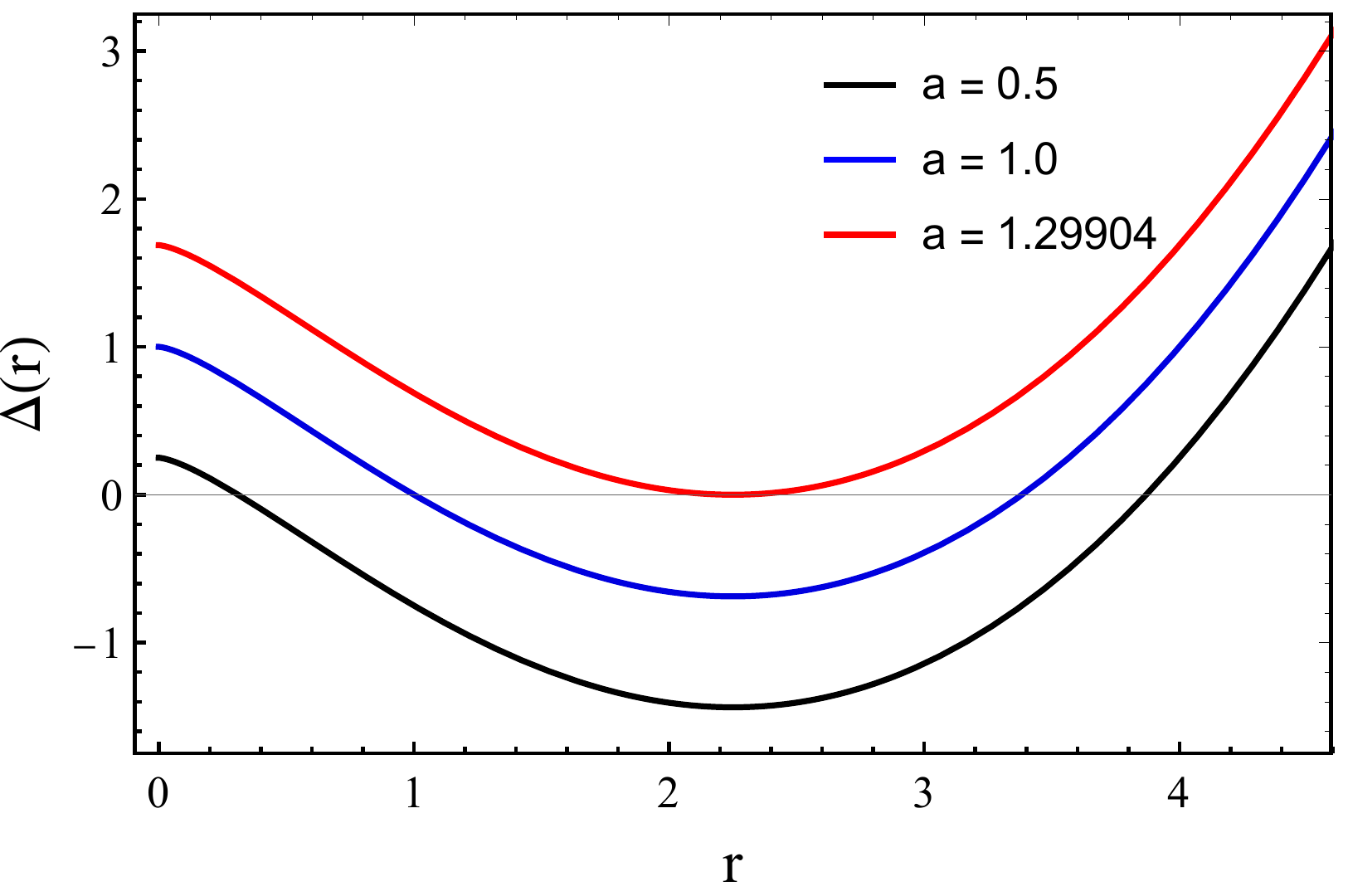}
 \includegraphics[width=0.45\textwidth]{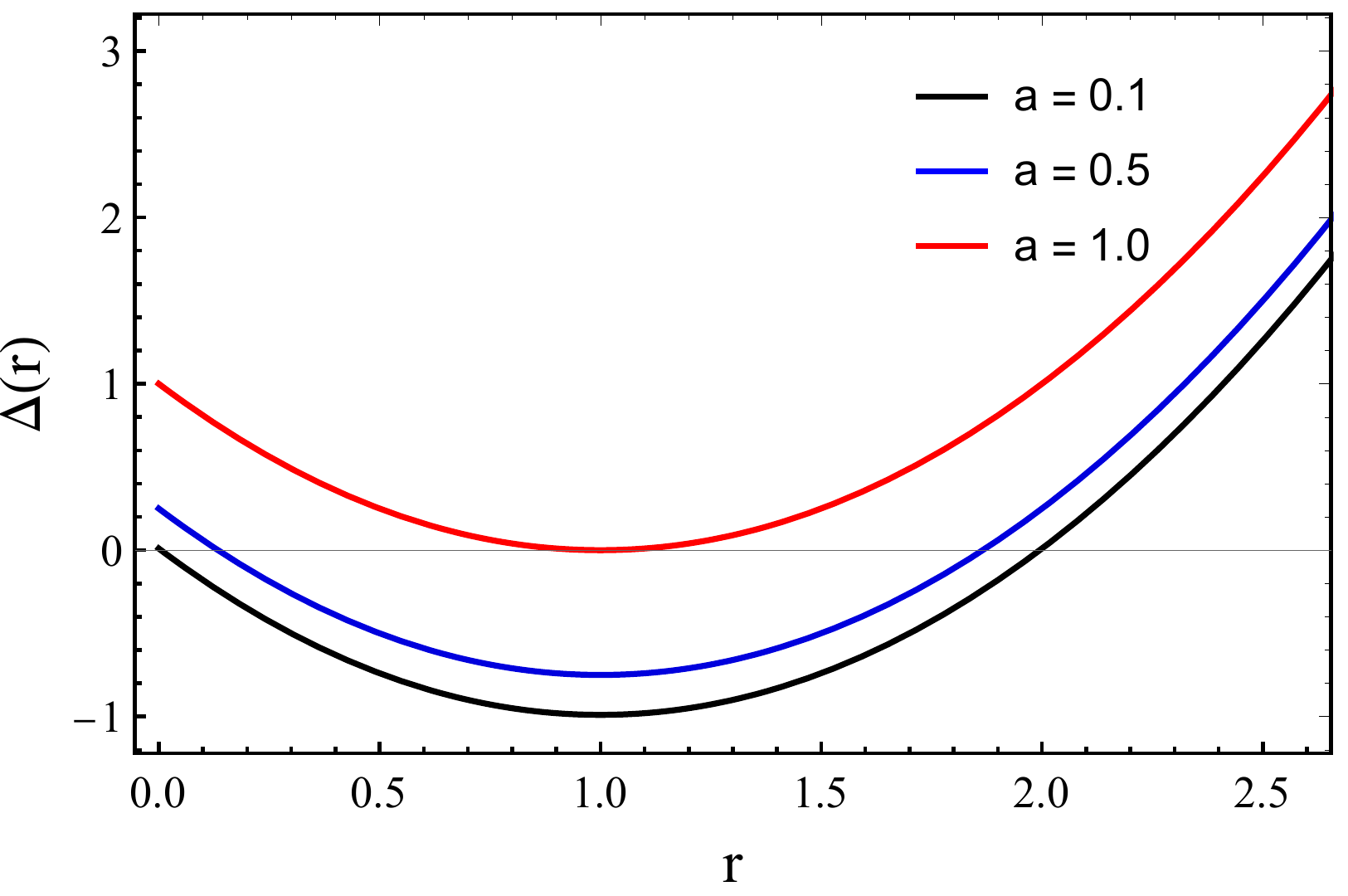}

\includegraphics[width=0.45\textwidth]{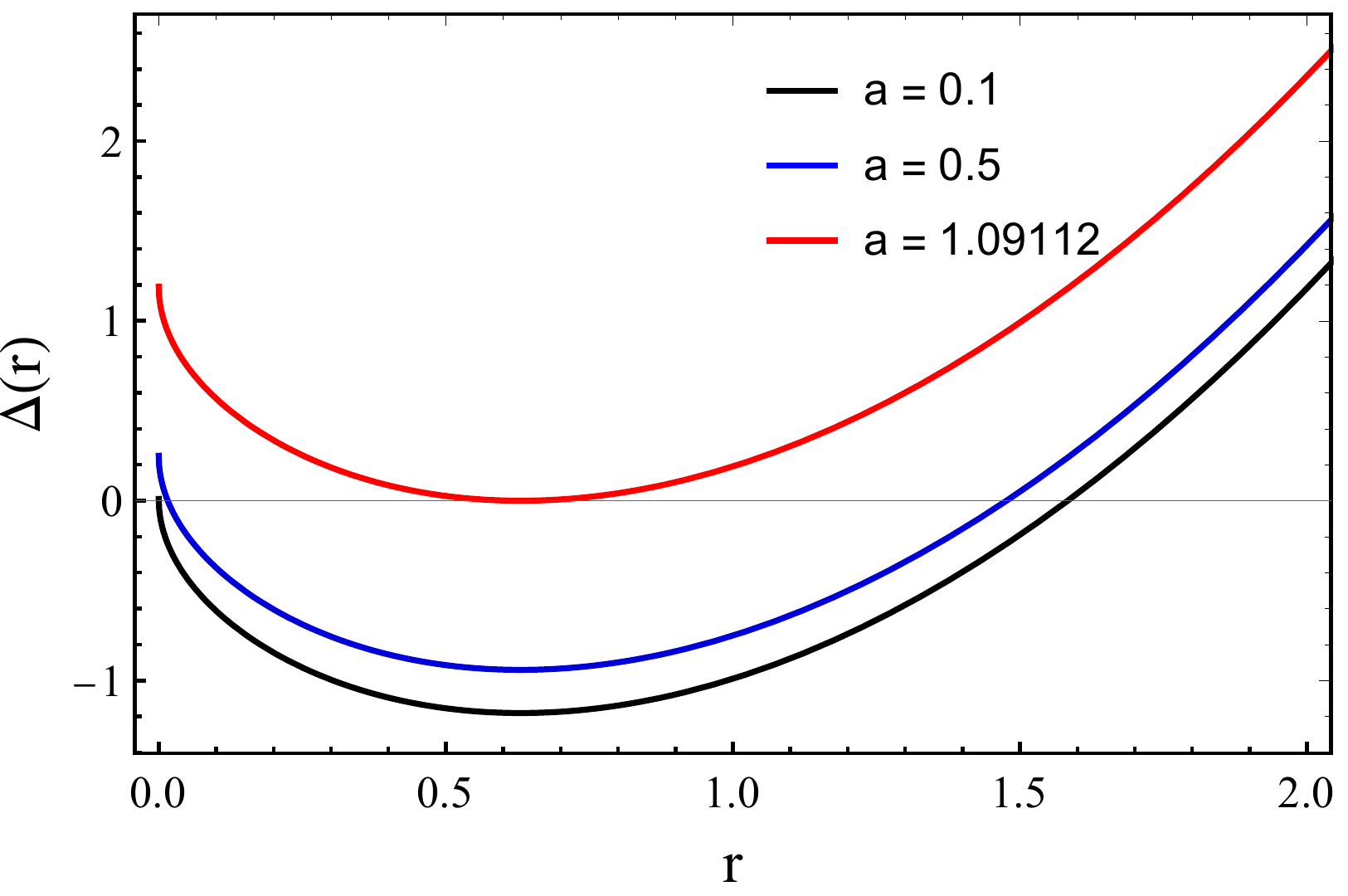}
\includegraphics[width=0.45\textwidth]{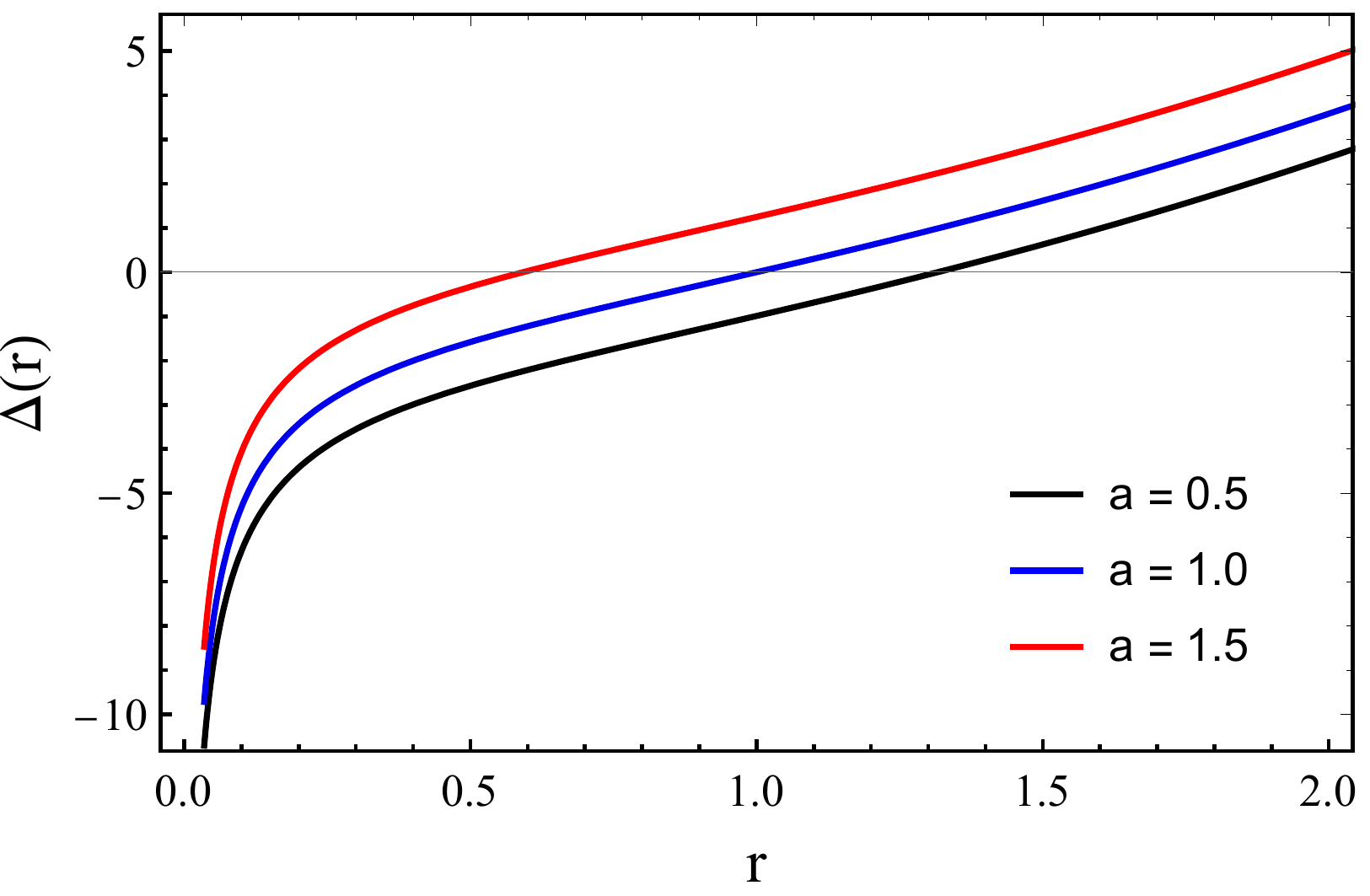}

 \caption{\label{fig:delta2} For $n=1$, $\Delta(r)$ is plotted in $D=6,7$ on the top and $D=8,9$ on the bottom.  }
\end{figure*}

Further we consider the ISCO for which one needs to solve the following standard equation
\begin{eqnarray}\label{isco3}
 \frac{\partial^2 V_{eff}(r)}{\partial r^2}\geq0\, ,
\end{eqnarray}
which determines radius of stable circular orbit. However, the ISCO radius is obtained on solving ${\partial^2 V_{eff}(r)}/{\partial r^2}=0$. It provides the minimum radius for which particles can move on stable circular orbits. The ISCO radius in Fig.~\ref{fig:dif} is located where the curves touch the horizontal axis. From Fig.~\ref{fig:dif} one can clearly see that the ISCO radius decreases up to $r_{ISCO} =2.25$ as given in Table~\ref{tab1} when black hole reaches its extremal rotation value.

\section{One or two horizons}\label{sec:one or two}

It turns out that if number of rotation parameters is less than the maximum allowed, $n = [(D-1)/2]$, in a given dimension, higher  dimensional Myers-Perry black holes can have only one horizon and thereby no extremal limit on rotation \cite{Shaymatov21a}. All this depends upon whether the horizon equation $\Delta = 0$, where $\Delta$ is in general given by Eq.~(\ref{Eq:D11}), has one or two positive roots. For it to have two positive roots, what is required is
\begin{eqnarray}\label{isco3}
 2n - \alpha > 0\, ,
\end{eqnarray}
where $\alpha = (D-2N-1)/N$. For $N=1$ Myers-Perry black holes, it becomes $2n+3-D > 0$; i.e. $2n+3 > D$. Recall that $n = [(D-1)/2]$ which implies $D = 2n+1, 2n + 2$ for all $n$ rotations being non-zero. So long as all rotation parameters are present, the inequality, $2n+3>D=2n+2$, would be trivially satisfied. However when one of rotations is zero, $n \to n-1$ in the same $D$ dimension, then it would be $2(n-1) +3 > 2n+1$ which is impossible. That is why whenever one of rotations is switched off, black hole would have only one horizon.

In the case of pure Lovelock rotating black hole we rewrite the above equation,
\begin{eqnarray}\label{isco3}
 2N(n+1) + 1 > D \, ,
\end{eqnarray}
which, for $n \to n-1$, would be always satisfied so long as $2Nn+1 > D$. That is for GB rotating black hole, $N=2$, for $n=1,2$ it would require $9 > D$ and $13>D$ respectively. For pure GB black hole, there would occur two horizons even for $(n-1)$ rotations; i.e. for single rotation in $D = 6, 7, 8$. This is in stark contrast to Myers-Perry class where black hole with one rotation has two horizons only in four dimension and none else. The above inequality is always violated for $N=1$ whenever number of rotations is $<n$ in a given dimension. Thus Myers-Perry black holes would, with one rotation being zero, have only one horizon while pure GB/Lovelock would continue to have two horizons as shown in Fig.~\ref{fig:delta2}.

It is rather interesting that even when black hole is rotating with $(n-1)$ rotations, Myers-Perry black hole has only one horizon. That alters its causal structure radically turning singularity from null to spacelike. A black hole has two horizons only when overall acceleration changes from the usual attraction to repulsion between the two horizons. This happens because of presence of rotation and charge as their gravitational contribution is repulsive opposing attraction due to mass. It turns out that the former could override the latter between two horizons so as to make overall acceleration change its sign from attraction to repulsion.

Note that horizons are given by positive roots of $\Delta = 0$. The first root (event horizon) is caused by dominance of attractive component while the second (Cauchy horizon) by dominance of repulsive component. It is therefore critical for repulsive component to dominate so as to cause the second (Cauchy) horizon. It turns out that whenever one of rotations is zero, this cannot happen for Myers-Perry black hole. To see why it cannot happen, note that $\Delta/r^{2n}$ serves as potential for acceleration and it is given by
\begin{eqnarray}\label{isco3}
 V = \Delta/r^{2n} &=& 1+ (a_1^2+a_2^2+...+a_n^2)/r^2 + ...\nonumber\\ &+& (a_1^2a_2^2...a_n^2)/r^{2n} - 2\mu/r^{D-3}\, .
\end{eqnarray}
In this expression clearly the last but one repulsive term would dominate when all rotations are non-zero. This is because $2n+3 > D \geq 2n+2$. On the other hand when one of rotations vanishes, it is the last attractive term which would continue to override because $2(n-1)+3 > D \geq 2n+1$ is violated. There would occur no change of sign in acceleration to cause the second (Cauchy) horizon.

\begin{figure*}
\centering
  \includegraphics[width=0.45\textwidth]{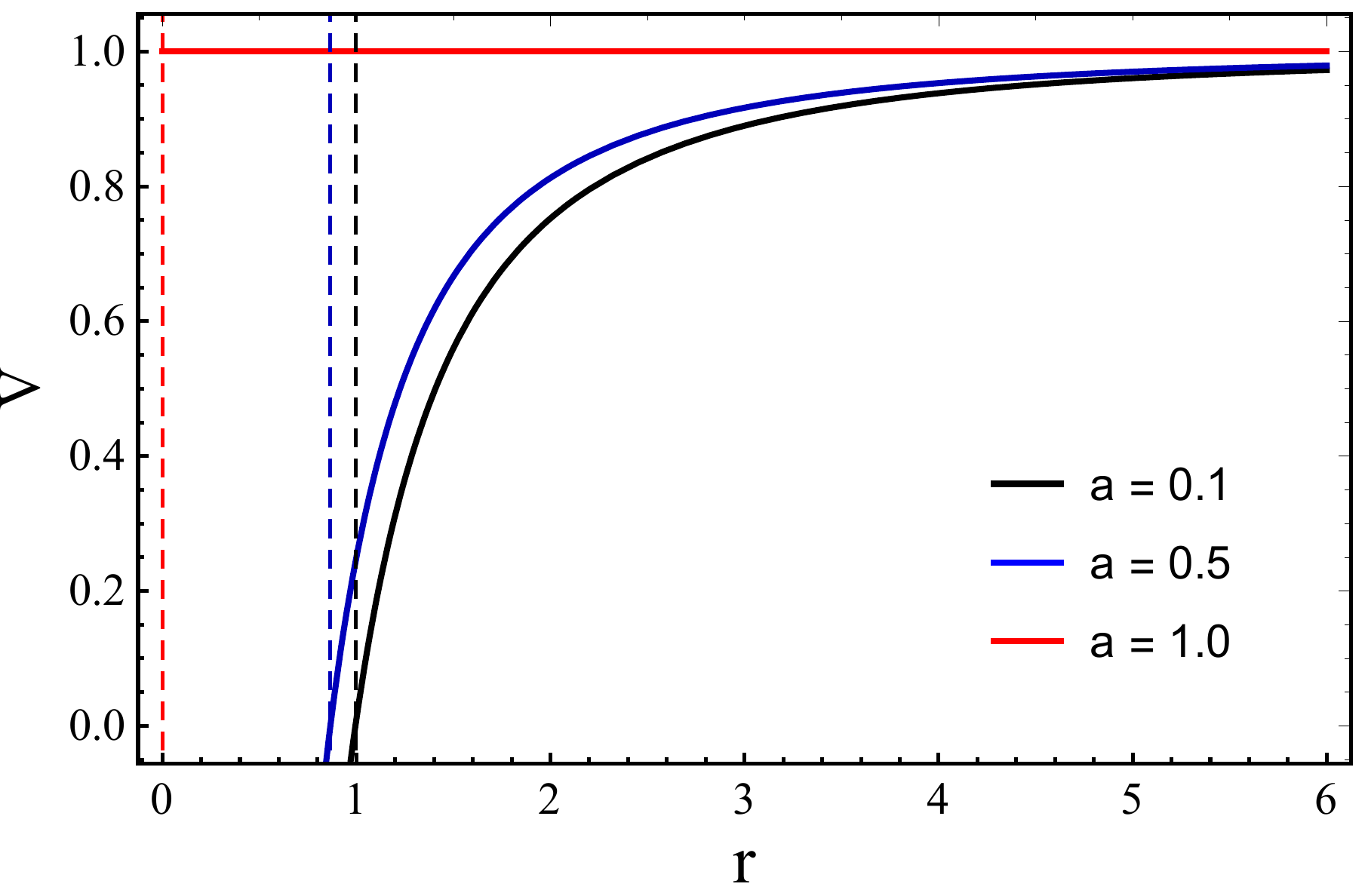}
  \includegraphics[width=0.45\textwidth]{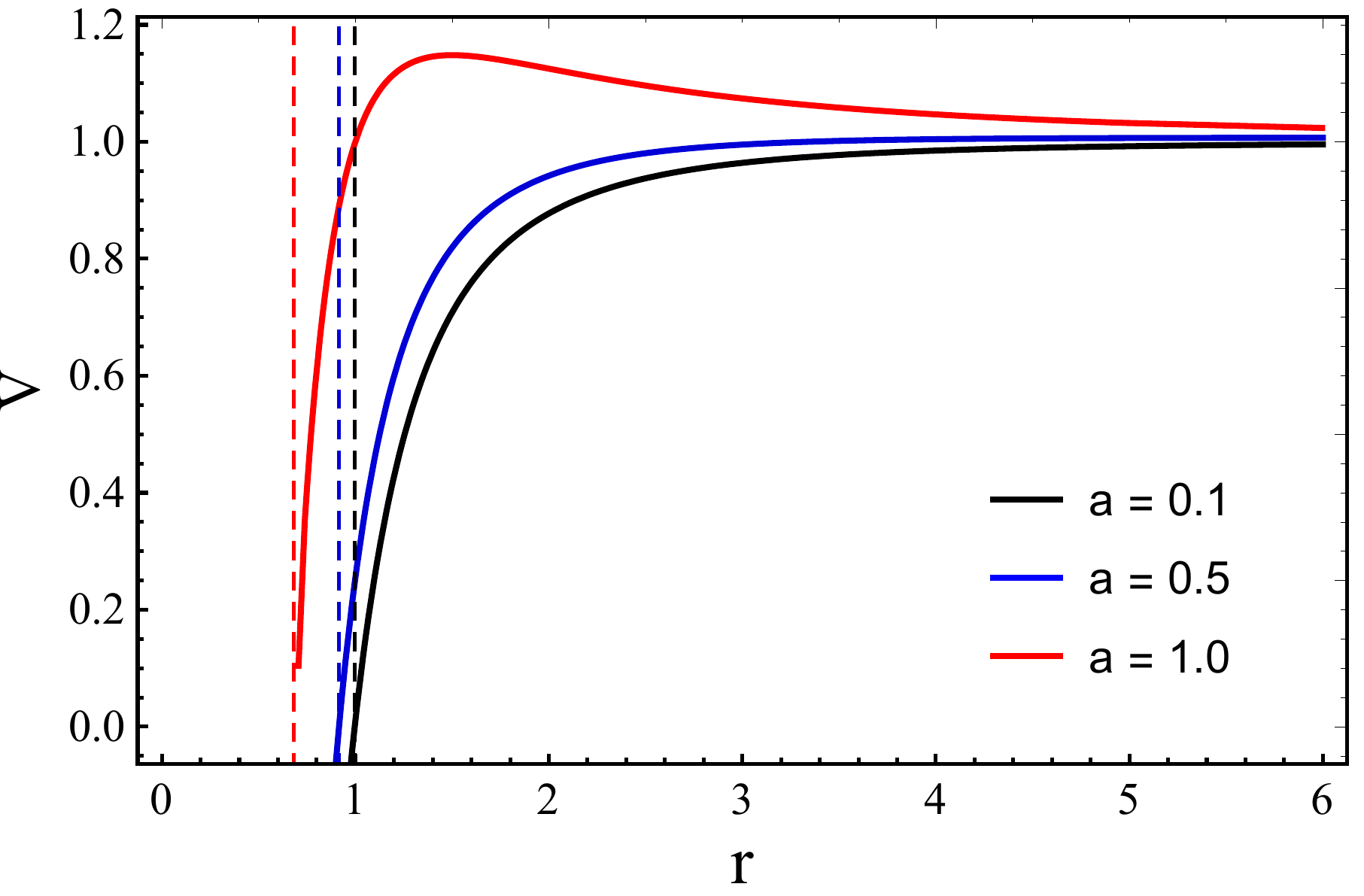}

  \includegraphics[width=0.45\textwidth]{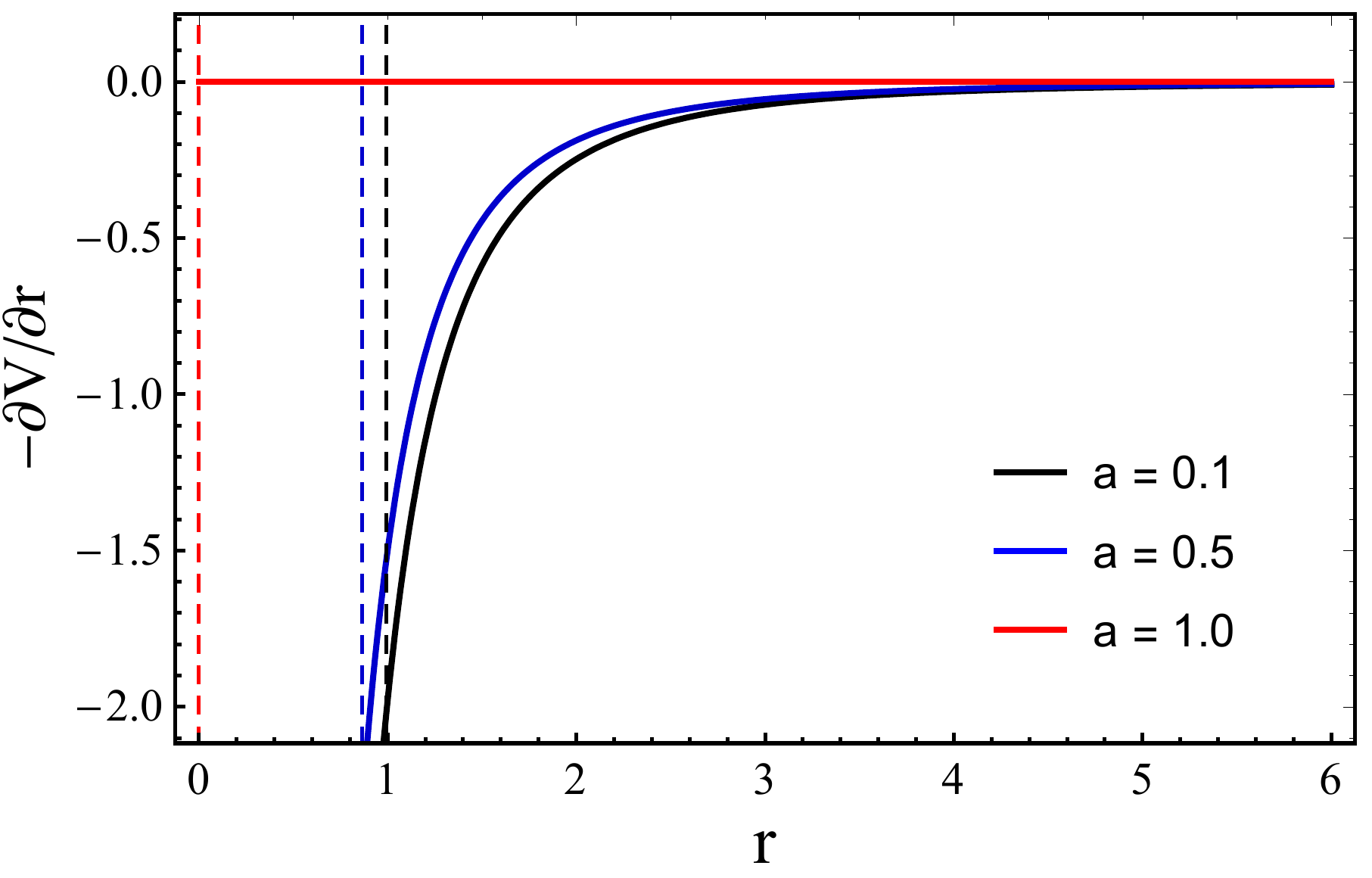}
  \includegraphics[width=0.45\textwidth]{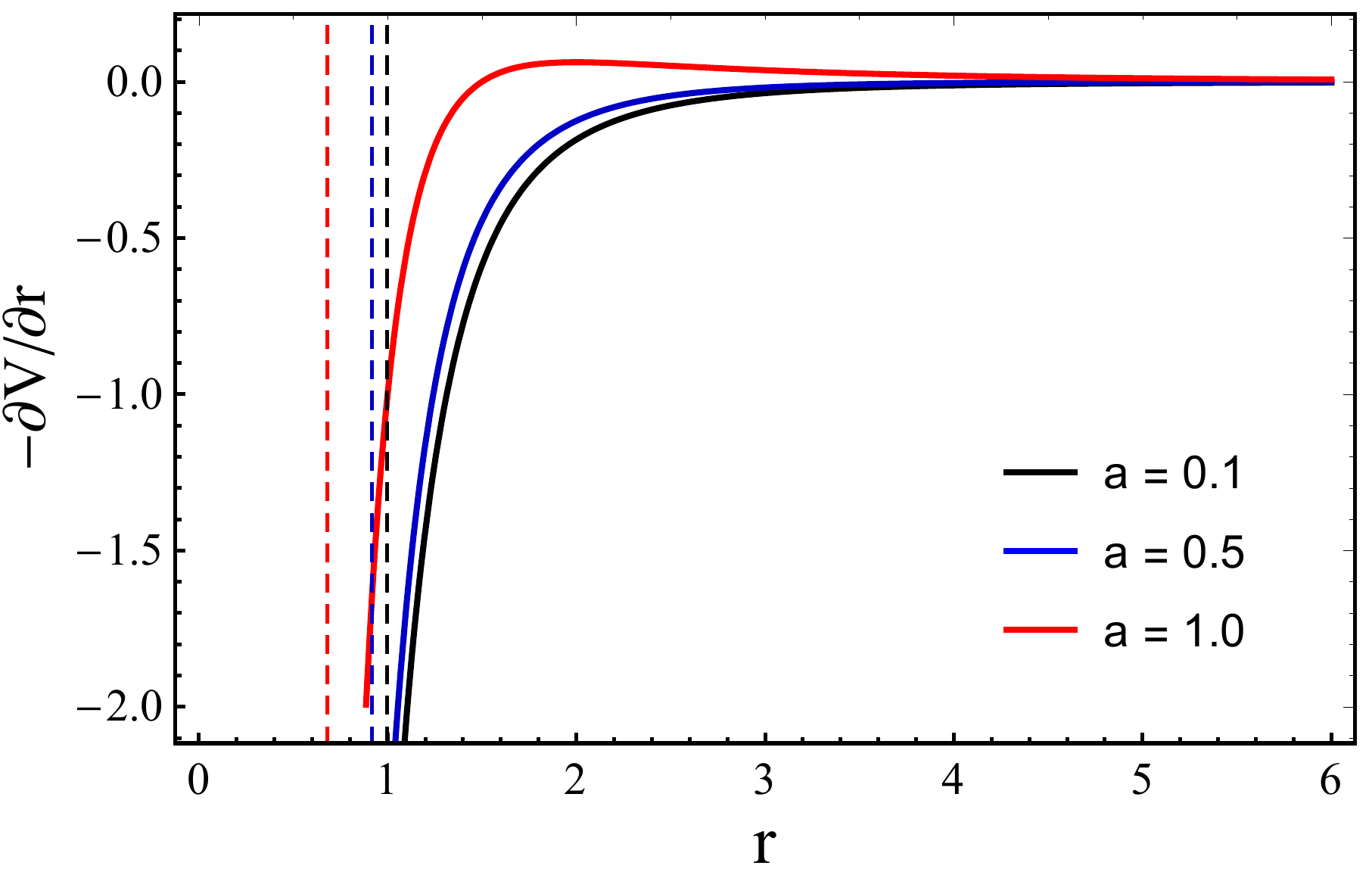}

 \caption{\label{fig:10} $ V$ and $-\partial V/\partial r$ for $n=1$ for Myers-Perry black hole in $D=5,6$ (left/right panels). The vertical dashed lines indicate location of horizon $r_h$.}
\end{figure*}

\begin{figure*}
\centering
  \includegraphics[width=0.45\textwidth]{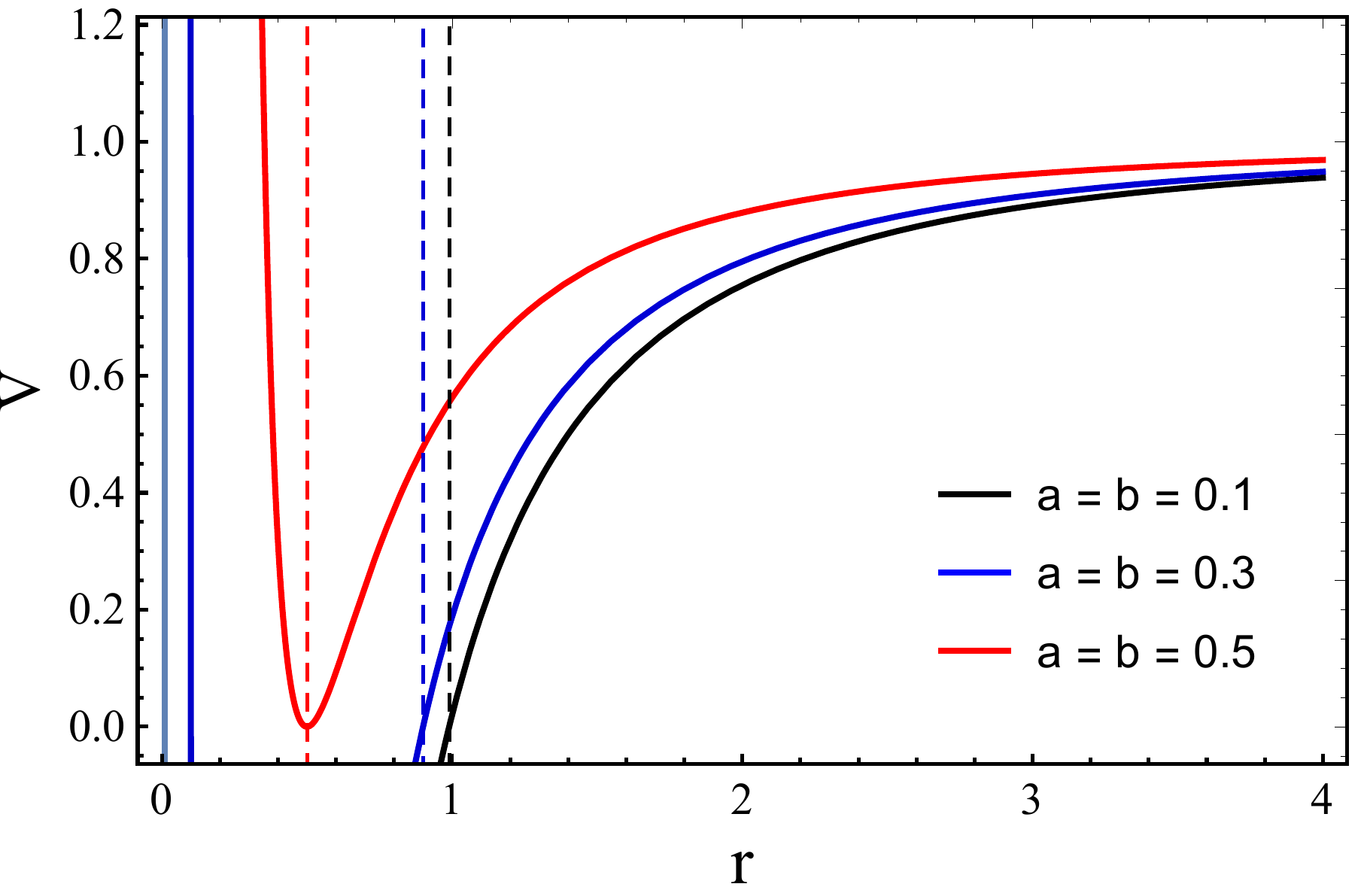}
  \includegraphics[width=0.45\textwidth]{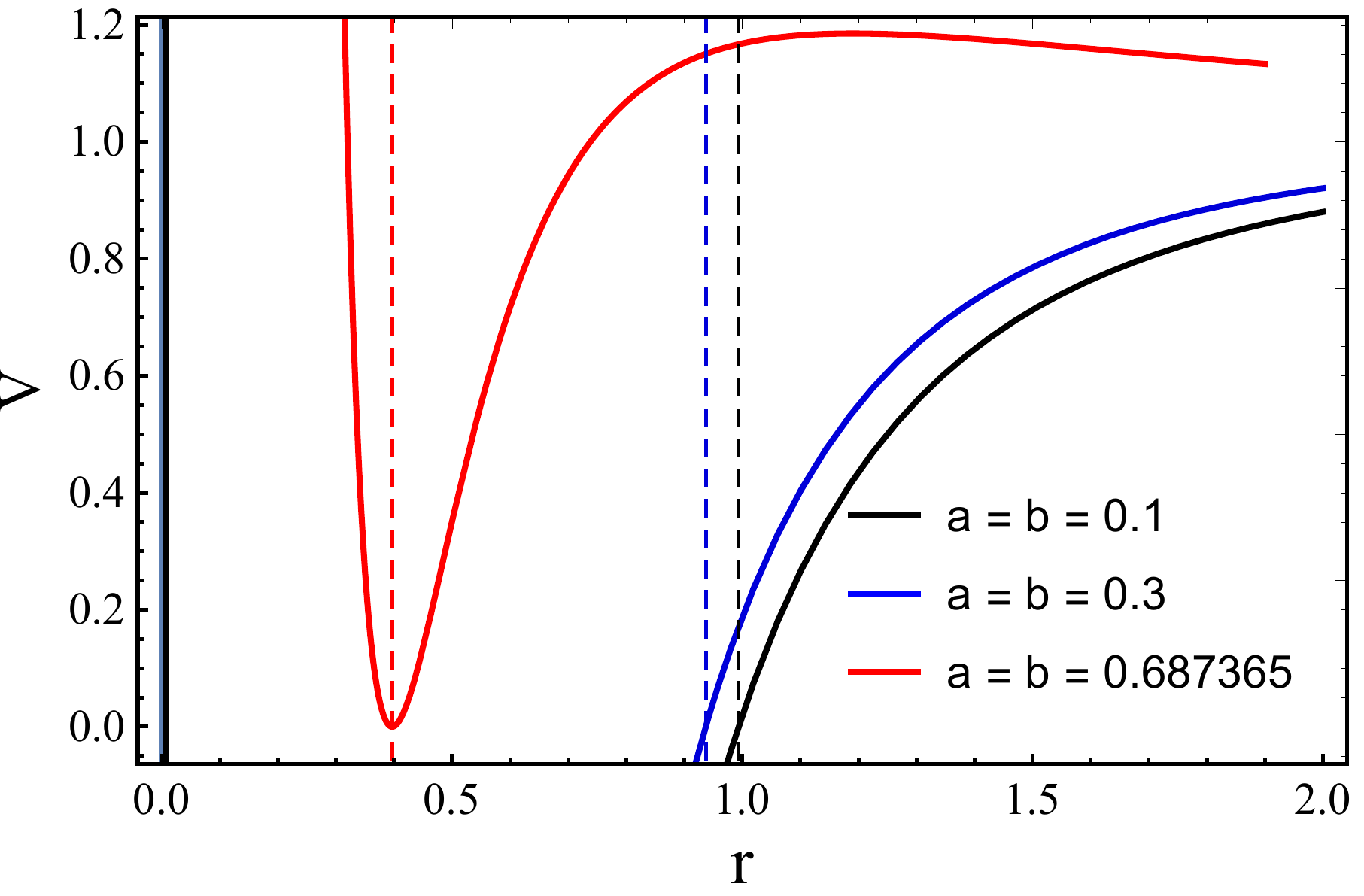}

  \includegraphics[width=0.45\textwidth]{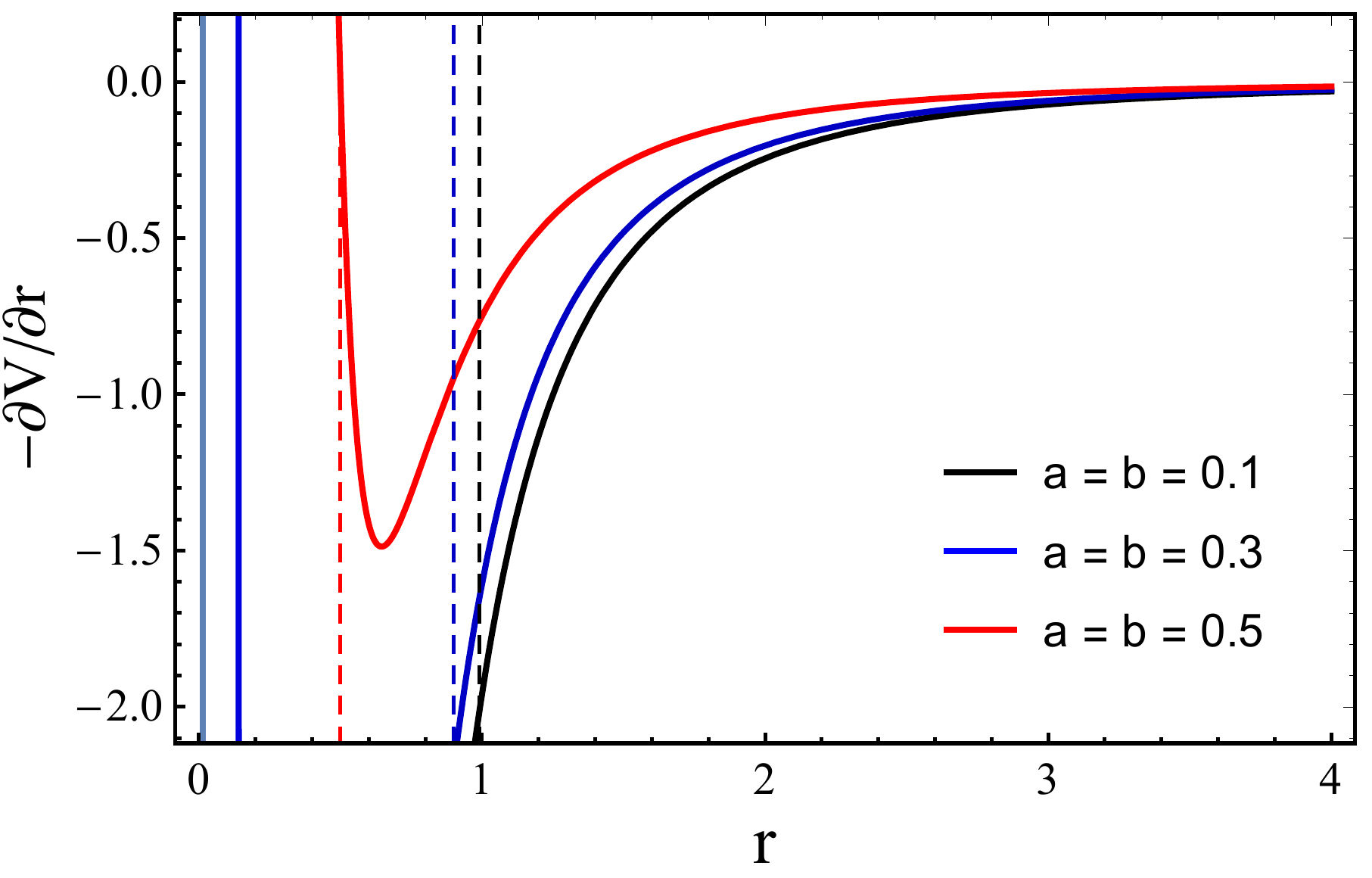}
  \includegraphics[width=0.45\textwidth]{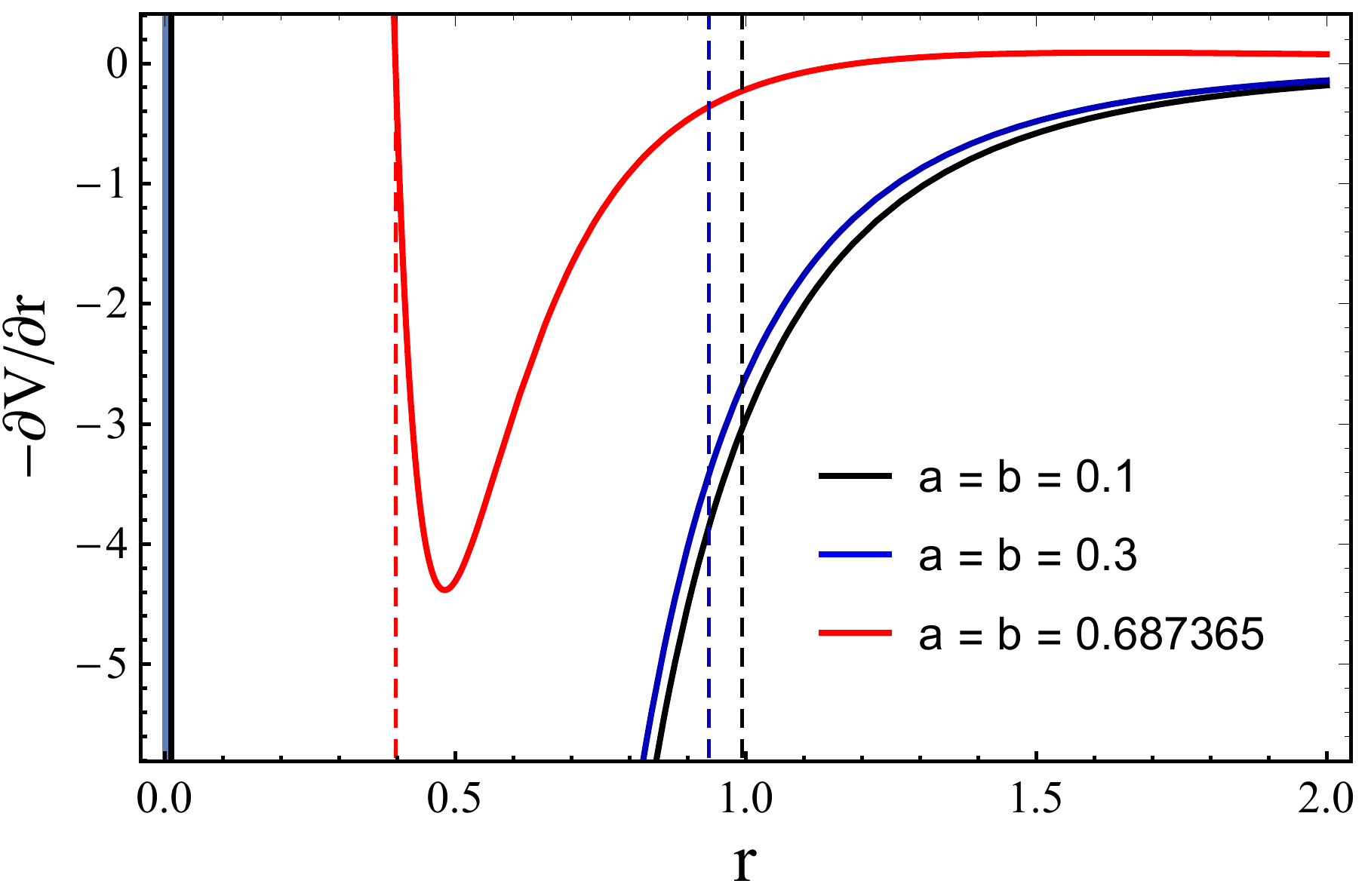}

 \caption{\label{fig:11}  $ V$ and $-\partial V/\partial r$ for $n=2$ for Myers-Perry black hole in $D=5,6$ (left/right panels). The vertical dashed lines indicate location of horizon $r_h$.}
\end{figure*}

\begin{figure*}
\centering
 \includegraphics[width=0.32\textwidth]{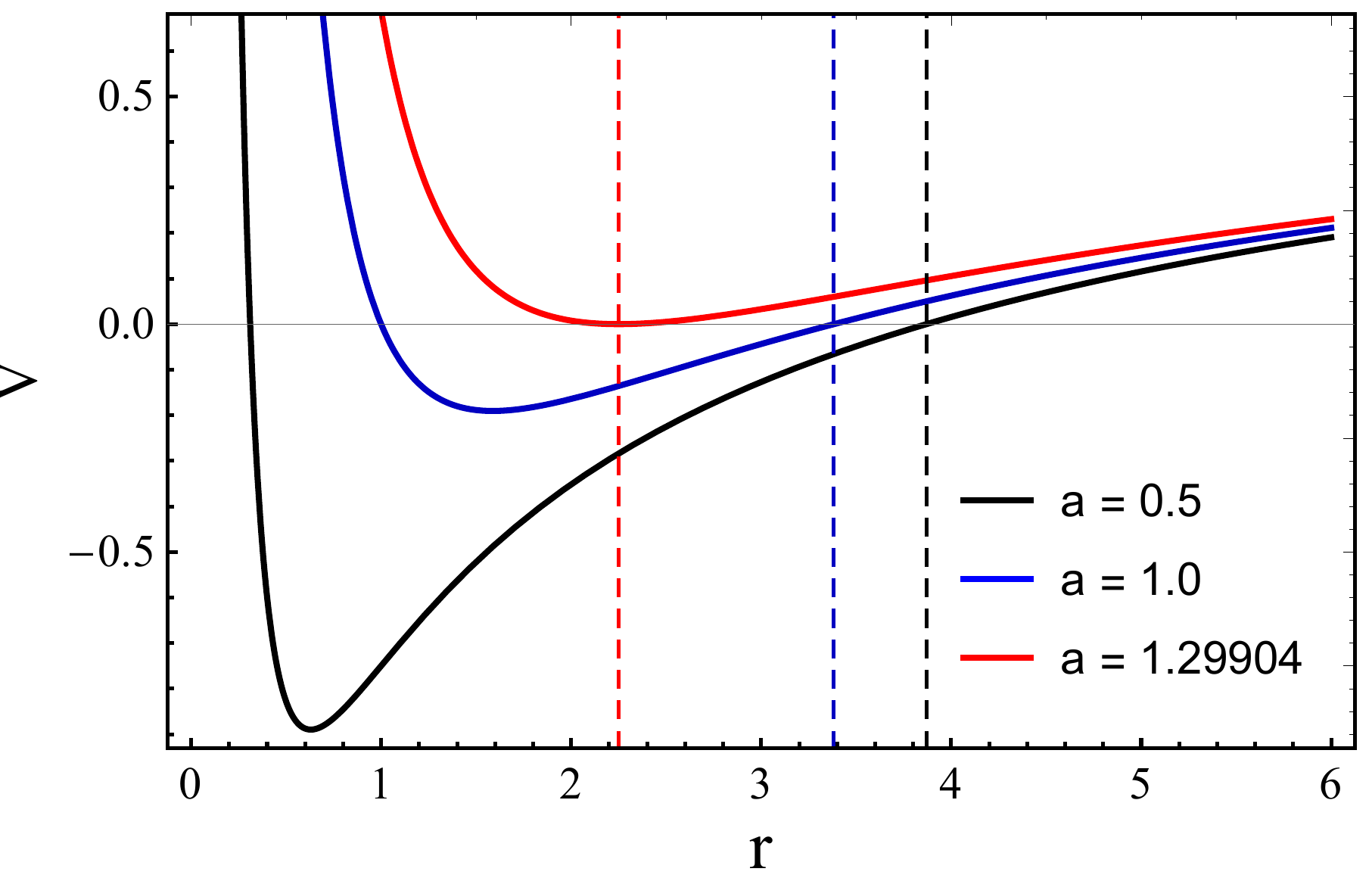}
 \includegraphics[width=0.32\textwidth]{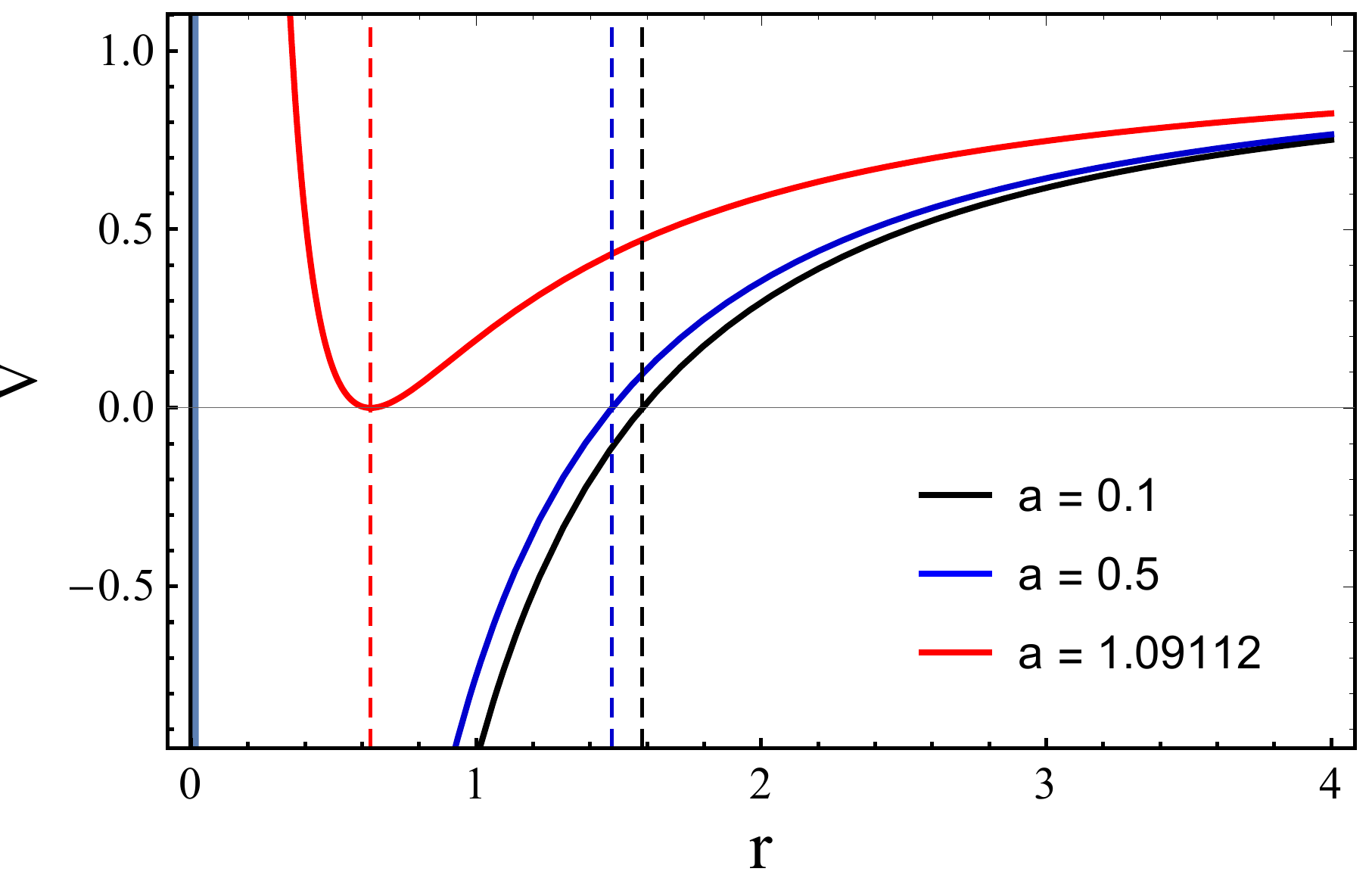}
\includegraphics[width=0.32\textwidth]{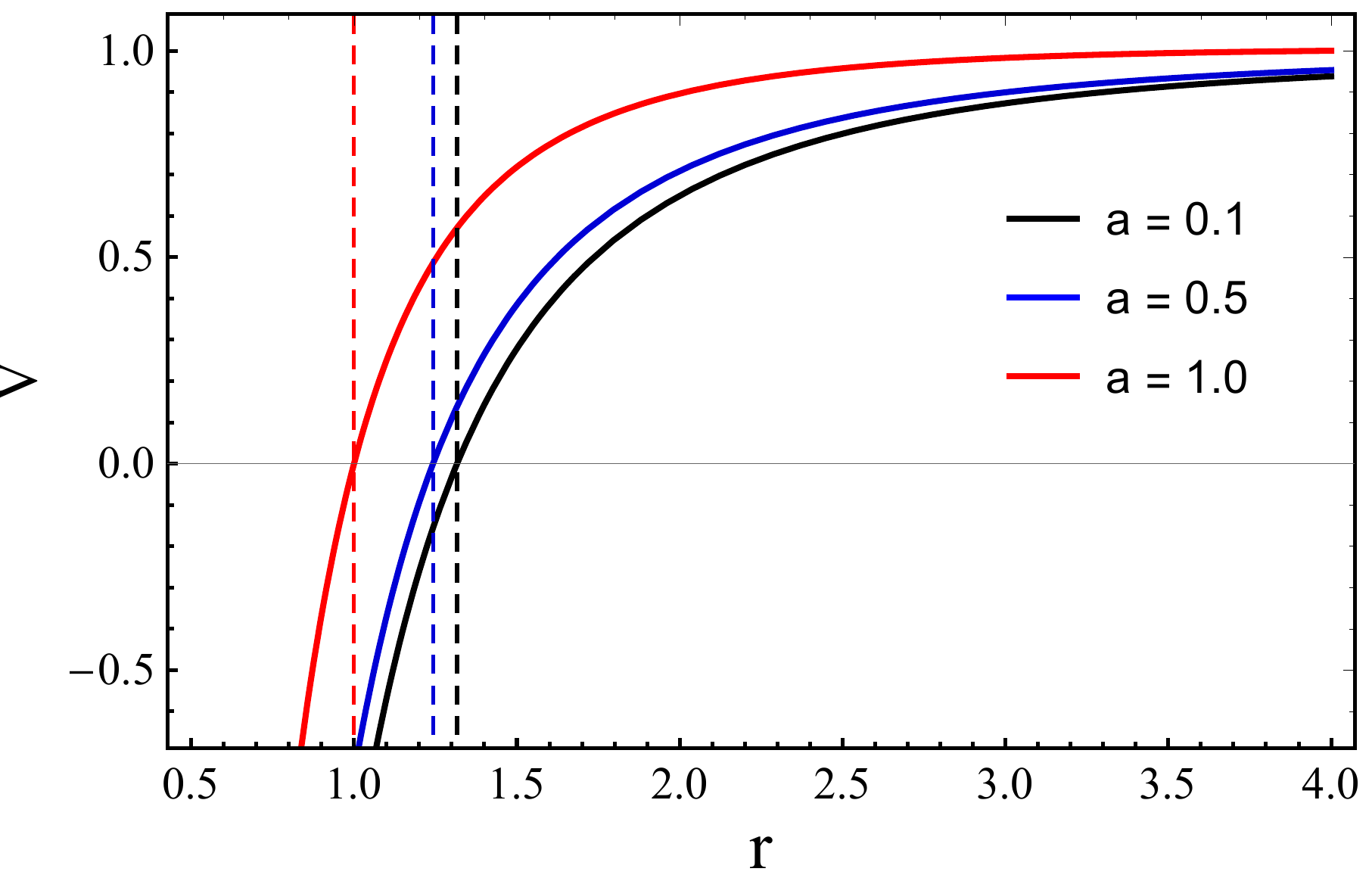}

\includegraphics[width=0.32\textwidth]{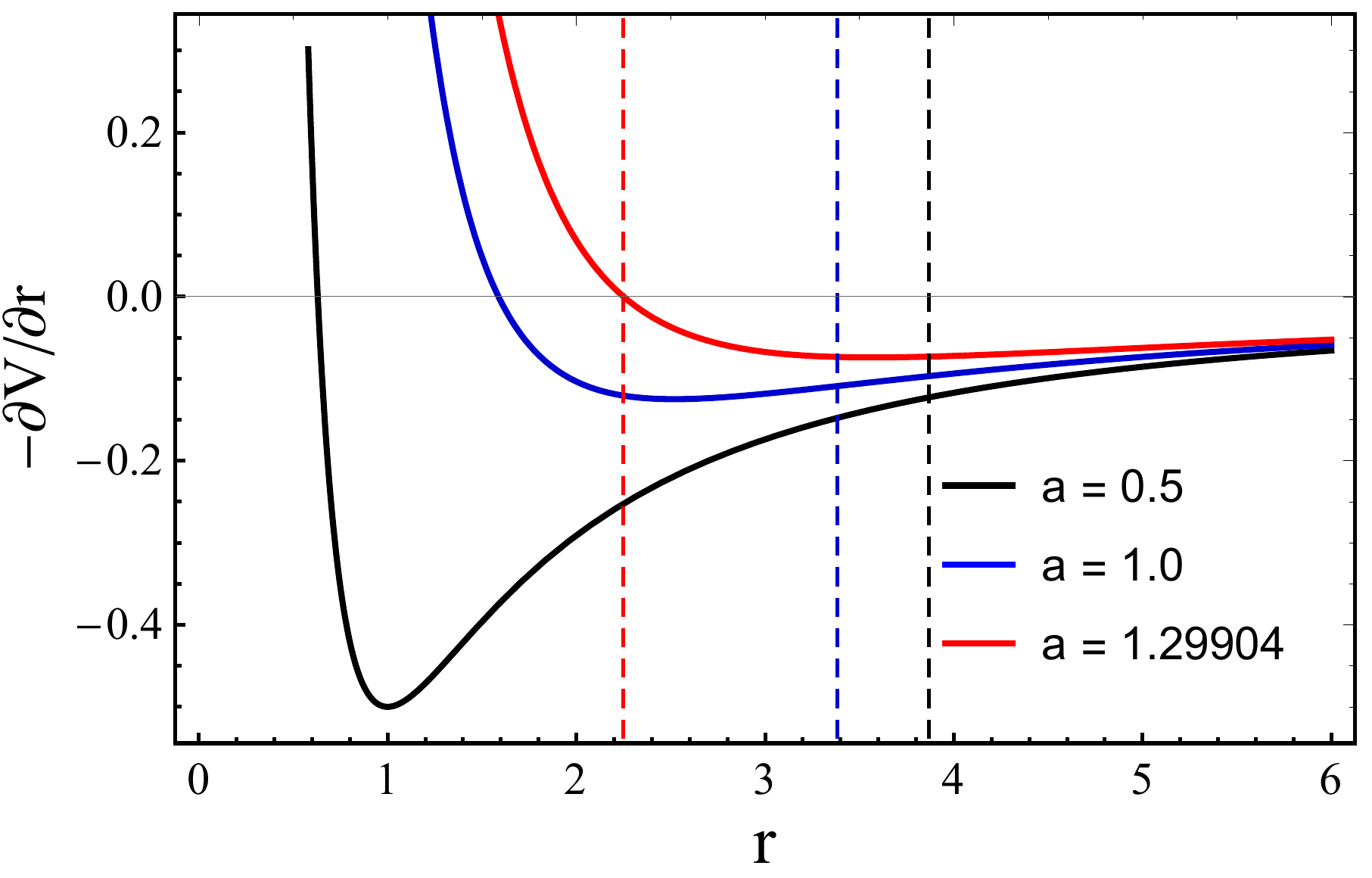}
 \includegraphics[width=0.32\textwidth]{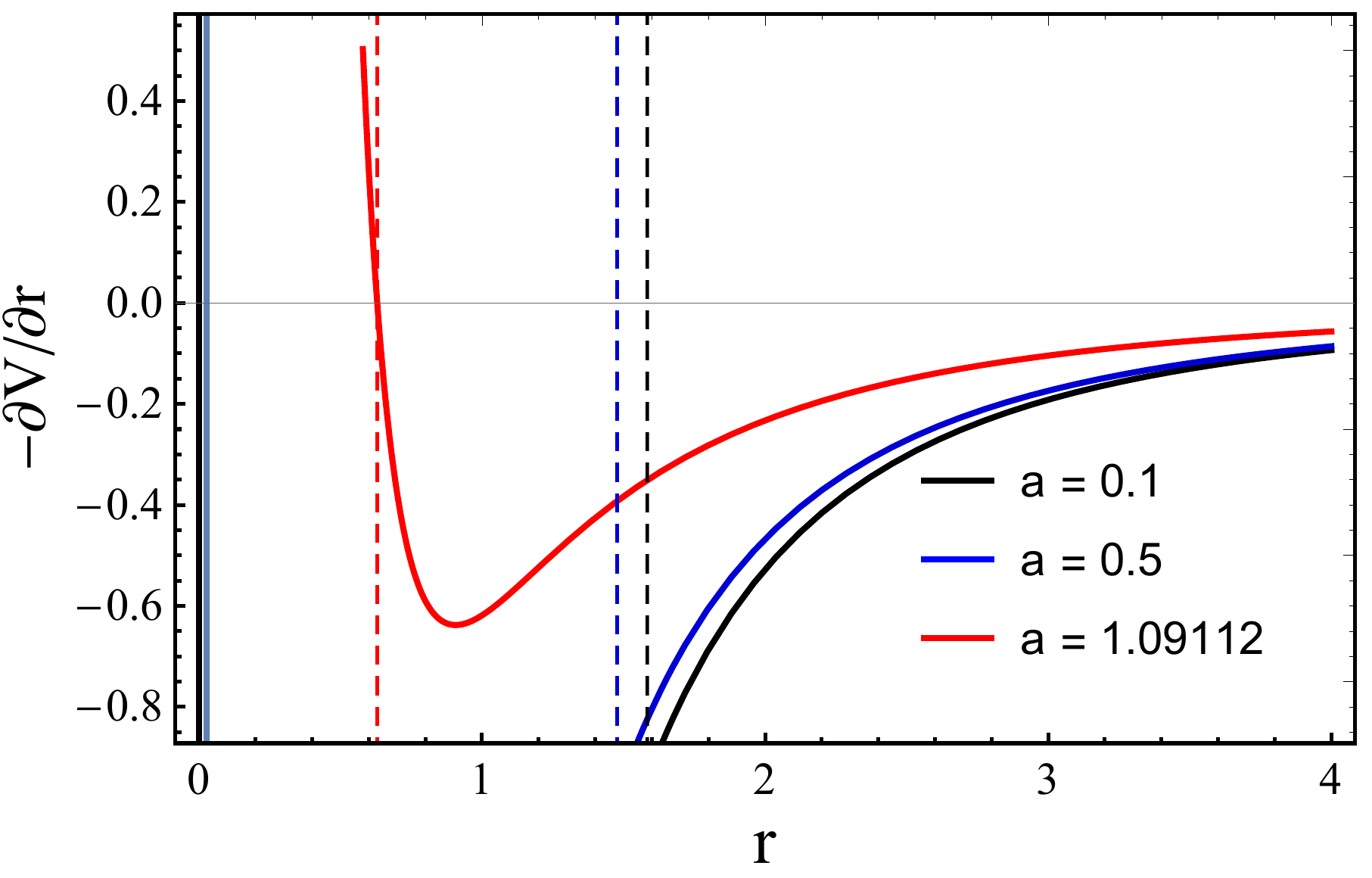}
\includegraphics[width=0.32\textwidth]{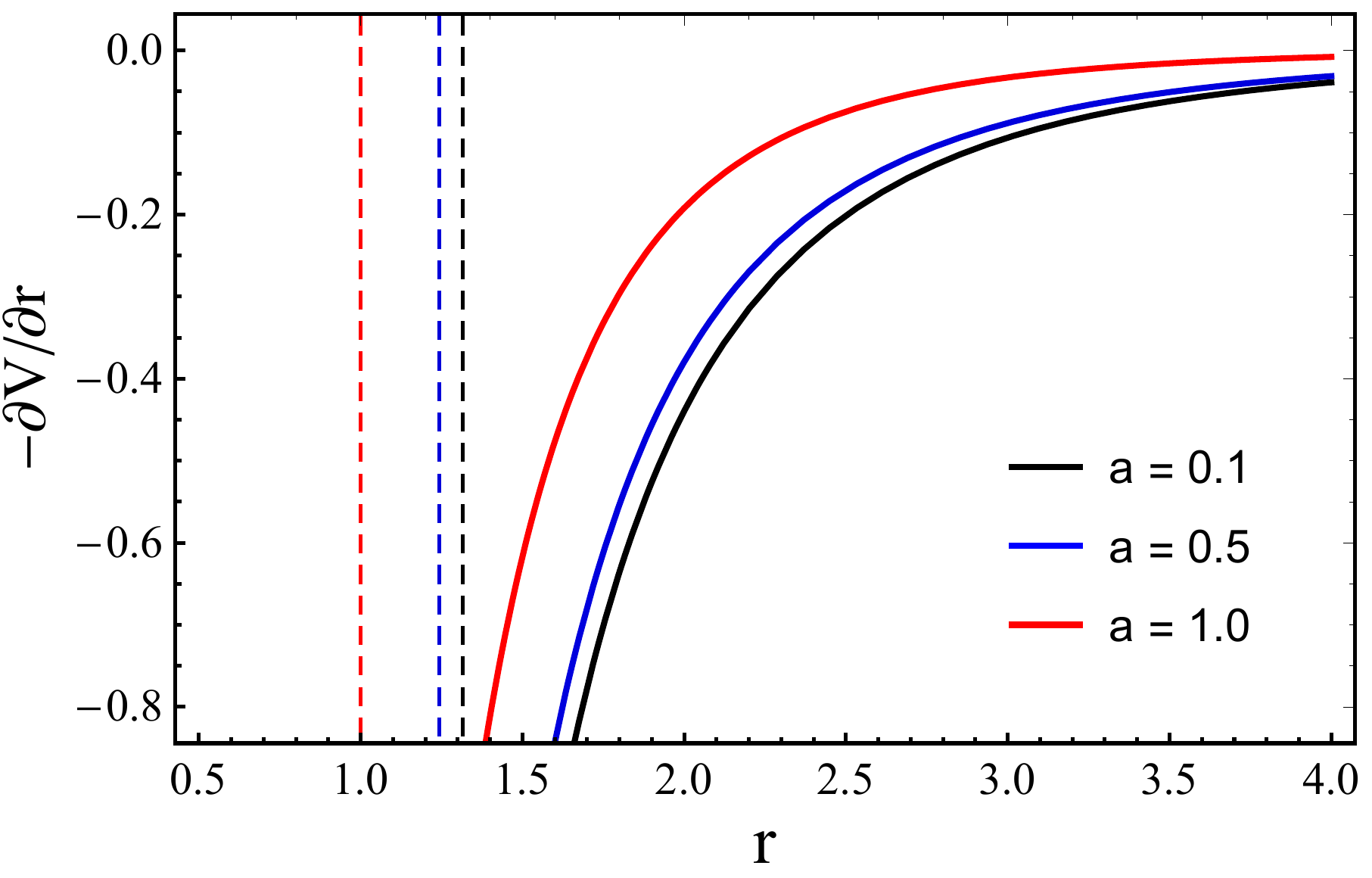}

 \caption{\label{fig:12} $ V$ and $-\partial V/\partial r$ are plotted for $n=1$ and $N=2$ in $D=6,8,9$ (left/middle/right panels). The vertical dashed lines indicate location of horizon $r_h$. }
\end{figure*}

\begin{figure*}
\centering
 \includegraphics[width=0.45\textwidth]{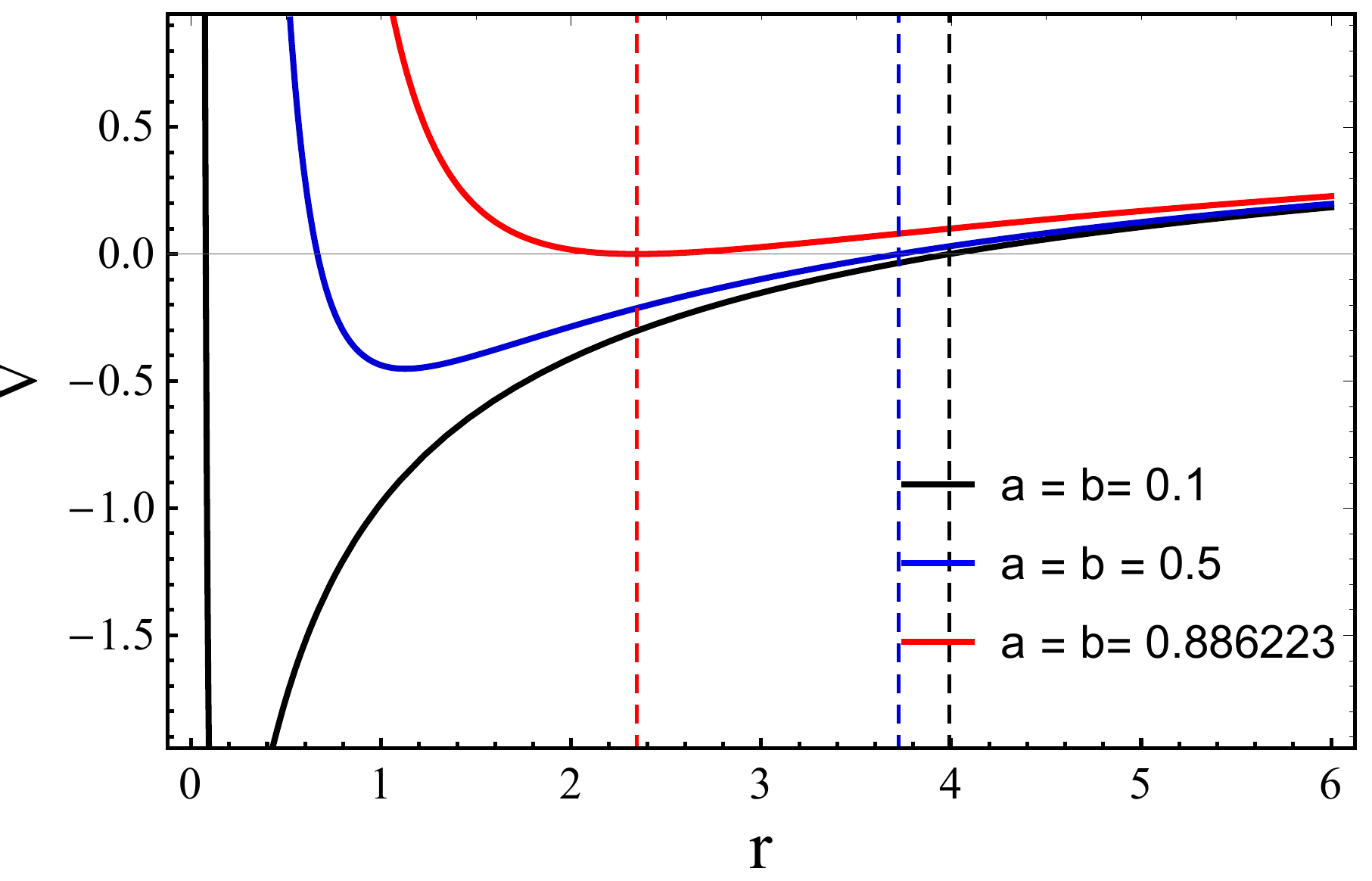}
 \includegraphics[width=0.45\textwidth]{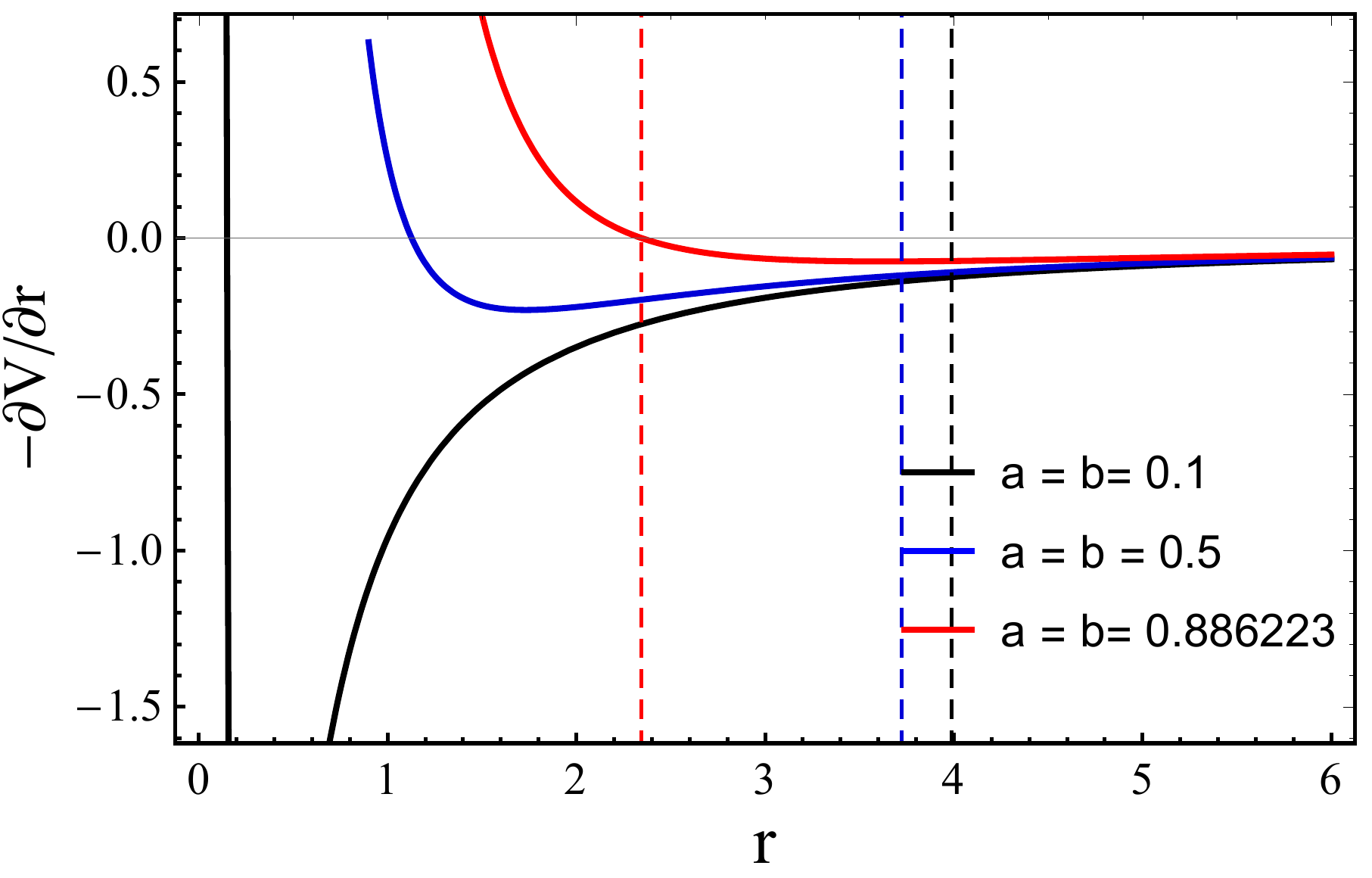}

 \caption{\label{fig:13} $ V$ and $-\partial V/\partial r$ are plotted for $n=2$ and $N=2$ in $D=6$. The vertical dashed lines indicate location of horizon $r_h$. }
\end{figure*}

All this what we have discussed above we wish to demonstrate by plotting $V$ and $-\partial V/\partial r$ in Figs. \ref{fig:10}$-$\ref{fig:13} for various cases. Figs.~\ref{fig:10} and \ref{fig:11} refer to Myers-Perry black hole in $D = 5, 6$ for $n=1, 2$ respectively. Note that in $D = 6$, $V$ approaches unity from above and therefore acceleration is repulsive for large $r$ which turns attractive closer to horizon. This is because asymptotically repulsive $a^2/r^2$ dominates over attractive $-\mu/r^3$ \cite{Dadhich20}. In $D =5$, however acceleration remains attractive all through. For $n = 1, 2$ respectively there occur one (Fig.~\ref{fig:10}) and two (Fig.~\ref{fig:11}) horizons. For acceleration there is no change of sign for the former while it is for the latter. Fig.~\ref{fig:12} refers to the case of $N=2$ pure GB black hole with single rotation in $D = 6, 8, 9$. In $D = 6, 7, 8$ ($D=7$ is not included in Fig.) there occurs change of sign for acceleration giving rise to two horizons. On the other hand for $D=9$, since there is no change of sign in acceleration, there occurs only one horizon. Finally Fig.~\ref{fig:13} refers to pure GB black hole with two rotations, exhibiting the same behavior as in Fig.~\ref{fig:12} for $D = 6, 8$.


\section{Discussion and Conclusion}\label{sec:Conclusion}

It has been known that bound orbits cannot occur for higher dimensional Schwarzschild black hole but they do occur for its pure Lovelock analogue in dimensions, $2N+2 \leq D \leq 4N$ \cite{Dadhich13}. The reason for that is that gravitational potential goes as $1/r^\alpha$ where $\alpha = (D-2N-1)/N$ while centrifugal potential always falls off as $1/r^2$. The latter is able to balance out the former for $N=1$, Einstein gravity only in $D=4$ and none else. However for  pure Lovelock black hole, they would always exist for $\alpha < 2$; i.e.,  $2N+2 \leq D \leq 4N$. For pure GB that would mean $D = 6, 7, 8$.

It is interesting that the same feature is carried forward for rotating black holes as well; i.e. no bound orbits around higher dimensional Myers-Perry black holes but they do exist for pure GB rotating black holes in dimensions, $D = 6, 7, 8$. In particular in seven dimension, effective potential for a single black hole rotation is exactly the same as that for the four dimensional Kerr black hole. For Myers-Perry black holes, effective potential has no minimum for potential well to form, and in particular for $D\geq 6$, $V_{eff} \geq 1$ always. Since there occurs no minimum, there can occur no stable circular orbits around higher dimensional Myers-Perry black holes. In contrast for pure GB rotating black holes, stable circular orbits, and thereby ISCO, would always occur in $D = 6, 7, 8$. In general this would be true for pure Lovelock rotating black holes in $2N+2 \leq D \leq 4N$. This is a nice discriminating property between Myers-Perry and pure GB/Lovelock rotating black holes.

So we have that ISCO would occur in general for $2N+2 \leq D \leq 4N$, which for $N=1$, Myers-Perry black hole gives only one $D=4$. Thus bound orbits can occur only for four dimensional Kerr black hole and none else in Myers-Perry class. In general there is no relation between occurrence of bound orbits and two horizons, however it turns out that whenever bound orbits exist, black hole has two horizons but the converse is not true. That is Myers-Perry black hole would always have two horizons so long as $2n+3 > D$ while it would have no bound orbits in all higher dimensions greater than four.

On the other hand for pure GB/Lovelock rotating black holes in dimensions $2N+2 \leq D \leq 4N$, there do exist two horizons even for a single rotation parameter and so does the extremal limit for rotation. The weak cosmic censorship conjecture (WCCC) could not be probed \cite{Shaymatov21a,Shaymatov20a} for Myers-Perry black holes with number of rotations being less than $n=[(D-1)/2]$, the maximum allowed for a given dimension $D$ because there exists no extremal limit. In contrast, for pure GB/Lovelock rotating black holes it could be probed even for single rotation parameter being non-zero \cite{Shaymatov20-pl}.

The existence of extremal limit for single rotation is yet another simple and interesting discriminator between Myers-Perry and pure GB/Lovelock rotating black holes. There is no relation between occurrence of number of horizons and bound orbits. The former requires $2N(n+1) + 1 > D$ while for the latter $D < 4N + 1$. This shows that for $N=1$ Myers-Perry black holes, there can occur no bound orbits in $D > 4$ while two horizons would always occur when all $n$ rotations are non-zero (Fig.~\ref{fig:11}). Whenever one of rotations is zero, there would occur only one horizon (Fig.~\ref{fig:10}). On the other hand for $N=2$ pure GB rotating black holes, two horizons would occur for $n = 1,2$ respectively in dimension windows, $6 \leq D \leq 8$ and $6 \leq D \leq 12$ while bound orbits would occur for $6 \leq D \leq 8$. That is, for $n=1$, existence windows for bound and two horizons coincide while for $n=2$, in the window, $9 \leq D \leq 12$, there occur two horizons but no bound orbits. For $D \geq 13$, there can only occur one horizon and of course no bound orbits.

In Newtonian theory a particle with non-zero angular momentum would always meet potential barrier that would prohibit its fall into the centre $r=0$. How does an astrophysical object like a star or black hole acquire angular momentum? The saving grace however is in the fact that a particle can go as close to the object as one likes and hit it with non-zero angular momentum. What is prohibited is that it cannot hit the center, $r=0$. In GR the situation is different, there occurs the stability threshold given by ISCO, below which no stable circular orbit can exist. The angular momentum of ISCO defines the minimum threshold for particle to have a stable circular orbit. That is, a particle with angular momentum less than the ISCO threshold would encounter no potential barrier and fall into the central object. That is how the central object or black hole could acquire angular momentum.

This raises an interesting question, for higher dimensional Myers-Perry black holes for which there can occur no ISCO, hence no particle with angular momentum can fall in. How do then higher dimensional rotating black holes form \cite{Dadhich20}? They cannot form by gravitational collapse or accretion (since there occur no ISCO which is critical for accretion disk to form for accretion process to ensue). On the other hand, ISCO does exist for pure GB/Lovelock rotating black hole in dimensions, $2N+2 \leq D \leq 4N$, defining the threshold value of angular momentum. And so accretion process could set leading to formation of rotating black hole. Thus by gravitational collapse/accretion only pure GB/Lovelock rotating black holes could be formed \cite{Dadhich20}.

{One may however ask how can one trust pure GB/Lovelock results which are based on the metric which is not an exact solution of the vacuum equation? It is a valid question. Even though it is not a solution yet the metric does describe a rotating black hole with all the expected features. More importantly the question of black hole formation depends upon existence of bound and consequently ISCO orbits. We know that bound orbits/ISCO do not exist around higher dimensional static black hole in Einstein gravity while they do for pure Lovelock gravity in $2N+2 \leq D \leq 4N$ dimensions~\cite{Dadhich13}. Since they do not exist for higher dimensional static black hole, they do not exist for the higher dimensional Myers-Perry rotating black hole as well. What emerges from this is the expectation that if they exist for static black hole, then only they do for rotating black hole. Since for pure GB/Lovelock theory they do exist for static black hole, they are therefore expected to exist for its rotating analogue as well. This is what we have shown that they do for the pure GB metric we have employed, even though it is not a solution of the vacuum equation. It however bears out the expected result that bound orbits/ISCOs do exist around pure GB rotating black hole. }

This was the main motivation for studying circular orbits around higher dimensional rotating black holes. The conclusion that follows is that higher dimensional rotating black holes in Einstein gravity  cannot be formed by gravitational collapse/accretion. They could however be formed only in pure GB/Lovelock gravity.

\section{Acknowledgments}
We warmly thank A. Aghababai and B. Mirza for sharing their work in Ref.~\cite{Amin-Mirza21}. ND acknowledges the support from the CAS
President’s International Fellowship Initiative Grant No.
2020VMA0014. SS acknowledges the support from Research Grant  No. F-FA-2021-432 of the Uzbekistan Ministry for Innovative Development. 

%
\section*{References}

\bibliographystyle{apsrev4-1}
\bibliography{gravreferences}

\end{document}